\documentclass{aa}  
\usepackage{graphicx}
\usepackage{txfonts}
\begin{document}

   \title{Advancing Machine Learning for Stellar Activity and Exoplanet Period Rotation}

   \subtitle{Insights from Kepler Light Curves}
   

   \author{Fatemeh Fazel Hesar\thanks{f.fazel.hesar@liacs.leidenuniv.nl}
          \inst{1,2,3}
          \and
          Bernard Foing\inst{2,3}\fnmsep
          \and
          Ana M. Heras \inst{4}\fnmsep
           \and
          Mojtaba Raouf\inst{2,3}\fnmsep          
          \and 
          Victoria Foing\inst{3}\fnmsep
          \and 
           Shima Javanmardi\inst{1}\fnmsep
           \and
           Fons J. Verbeek\inst{1}\fnmsep}

   \institute{Leiden Institute of Advanced Computer Science (LIACS), Leiden University, Einsteinweg 55, 2333 CC Leiden, The Netherlands\\
              \email{f.fazel.hesar@liacs.leidenuniv.nl}
        \and
        Leiden Observatory, Leiden University, P.O. Box 9513, 2300 RA Leiden, Netherlands
        \and
        LUNEX \& Eurospacehub-Green academy \&, ESA BIC, Noordwijk, Netherlands
        \and
        Directorate of Science, European Space Research and Technology Center (ESA-ESTEC), Keplerlaan 1, NL-2201 AZ, Noordwijk, Netherlands
               }
   \date{Received Aug 29, 2024; accepted September, 2024}
  \abstract
   {\begin{abstract}
    
   This study applied machine learning models to estimate stellar rotation periods from corrected light curve data obtained by the NASA Kepler mission. Traditional methods often struggle to estimate rotation periods accurately due to noise and variability in the light curve data. The workflow involved using initial period estimates from the LS-Periodogram and Transit Least Squares techniques, followed by splitting the data into training, validation, and testing sets. We employed several machine learning algorithms, including Decision Tree, Random Forest, K-Nearest Neighbors, and Gradient Boosting, and also utilized a Voting Ensemble approach to improve prediction accuracy and robustness.
   The analysis included data from multiple Kepler IDs, providing detailed metrics on orbital periods and planet radii. Performance evaluation showed that the Voting Ensemble model yielded the most accurate results, with an RMSE approximately 50\% lower than the Decision Tree model and 17\% better than the K-Nearest Neighbors model. The Random Forest model performed comparably to the Voting Ensemble, indicating high accuracy. In contrast, the Gradient Boosting model exhibited a worse RMSE compared to the other approaches. Comparisons of the predicted rotation periods to the photometric reference periods showed close alignment, suggesting the machine learning models achieved high prediction accuracy. The results indicate that machine learning, particularly ensemble methods, can effectively solve the problem of accurately estimating stellar rotation periods, with significant implications for advancing the study of exoplanets and stellar astrophysics.

\end{abstract}}


   \maketitle

\section{Introduction}

Accurately determining stellar rotation periods from photometric light curve data is essential in studying exoplanets and stellar astrophysics. Stellar rotation periods reveal important information about the physical properties and evolutionary paths of stars, which is important for characterizing exoplanetary systems \citep{aigrain2015testing, reinhold2015detecting}. Applied machine learning models, including Random Forest and Gradient Boosting, have been used to estimate rotation periods from Kepler light curves, demonstrating the potential of these methods. In addition high-resolution spectroscopy is an effective technique for estimating the rotation periods of late-type stars, as demonstrated in earlier studies \citep{Char1993}. 

The study of exoplanets, particularly the detection and characterization of their transits, has emerged as a cornerstone in modern astrophysics. Exoplanet transits, the periodic dimming of a star's brightness as a planet passes in front of it, offer a unique opportunity to probe the physical properties, composition, and potential habitability of distant worlds beyond our solar system \citep{Seager2003}.

The Kepler Space Mission, launched in 2009, revolutionized exoplanet research by continuously monitoring stellar brightness and detecting thousands of exoplanet candidates \citep{Borucki2010}. This extensive data has enabled detailed studies of planetary sizes, orbital periods, and atmospheric conditions, significantly advancing our understanding of exoplanetary systems \citep{Mullally2015}. With long-term accurate space photometry from missions such as COROT, Kepler, and TESS, it is possible to monitor sunspots, rotation periods, and differential rotation. For example, COROT data has been utilized in studies by \cite{Lanza2009A, Lanza2009B}. 

Accurately characterizing exoplanet transits from Kepler light curves is challenging due to various noise sources, instrumental artifacts, and stellar variability \citep{Gilliland2010}. Traditional analysis methods often struggle to distinguish the subtle signals of exoplanet transits from these confounding factors, limiting our ability to extract precise transit parameters and understand the underlying planetary systems. Recent studies have shown that machine learning (ML) models are effective in distinguishing exoplanet signals from noise \citep{Armstrong2021, LeCun2015}. 

ML techniques, such as Random Forest (RF), k-nearest Neighbors (KNN), Decision Tree (DT), and Gradient Boosting (GB), have demonstrated remarkable success in modeling instrumental and astrophysical noise in light curves \citep{Shallue2018}. These models were chosen for their ability to handle complex, non-linear relationships in large datasets, which is essential for accurately characterizing exoplanet transits amidst significant noise and variability. Random Forest is known for its robustness and ability to handle high-dimensional data without overfitting \citep{Breiman2001}. Decision Tree provides interpretability and simplicity \citep{Quinlan1986}, while k-nearest Neighbors offers a non-parametric approach to classification that can capture local data patterns \citep{Cover1967}. Gradient Boosting is effective in improving model accuracy by combining weak learners to form a strong predictive model \citep{Friedman2001}. To optimize the performance of these models, we employed hyperparameter tuning techniques such as grid search and random search. Grid search involves systematically searching through a manually specified subset of the hyperparameter space, while random search samples a fixed number of hyperparameter combinations from a specified distribution. These optimization techniques were important in identifying the best-performing parameters for each model, ensuring that they could accurately identify and characterize exoplanet transits from the noisy light curve data.

In addition to Kepler, current missions such as the Transiting Exoplanet Survey Satellite (TESS), the James Webb Space Telescope (JWST), and in the early future the European Space Agency's PLATO mission, provide even more precise data for exoplanet detection and characterization. These missions will further refine our understanding of exoplanetary systems and the mechanisms driving stellar and planetary interactions. A recent study presents a machine learning-based method using a Gaussian Process Classifier to validate 50 new exoplanets from Kepler data, highlighting the effectiveness and adaptability of this approach for missions like TESS \citep{Armstrong2021}.

In this study, we utilize the power of ML models to enhance the prediction of rotation periods in Kepler light curves. Our approach involves employing a range of ML algorithms, including RF, KNN, DT, and GB, to develop robust models capable of accurately identifying and characterizing exoplanet transits. Additionally, we introduce a novel Best-Model (BM) approach that combines multiple ML techniques to provide a more comprehensive analysis of exoplanet transits. To select our best model, we employ an ensemble voting mechanism that integrates the strengths of these diverse algorithms, providing a holistic and accurate rotation period analysis. The primary objective of our research is to develop an ML model. These parameters, including the transit's duration, depth, and shape, provide information about the size, orbital period, and potential atmosphere of exoplanets \citep{Winn2010}. We place specific emphasis on accurately determining the rotation period of the host star, as it influences the observed transit signals and helps distinguish them from stellar activity \citep{McQuillan2013}. This comparative analysis allows us to assess the strengths and limitations of the ML models in accurately determining the stellar rotation period and highlights their potential applications in future exoplanet studies.

This study is organized into several key sections. Section \ref{sec:data} describes the data used in the analysis. The methodology, including data processing and the ML models employed, is detailed in Section 4. Section 5 discusses the results and predictions obtained through the study. We then test our approach on the Kepler star dataset in Section 6. Finally, Section 7 provides a summary of the overall findings.

   \begin{figure*}
   \centering
   \includegraphics[width=0.8\linewidth]{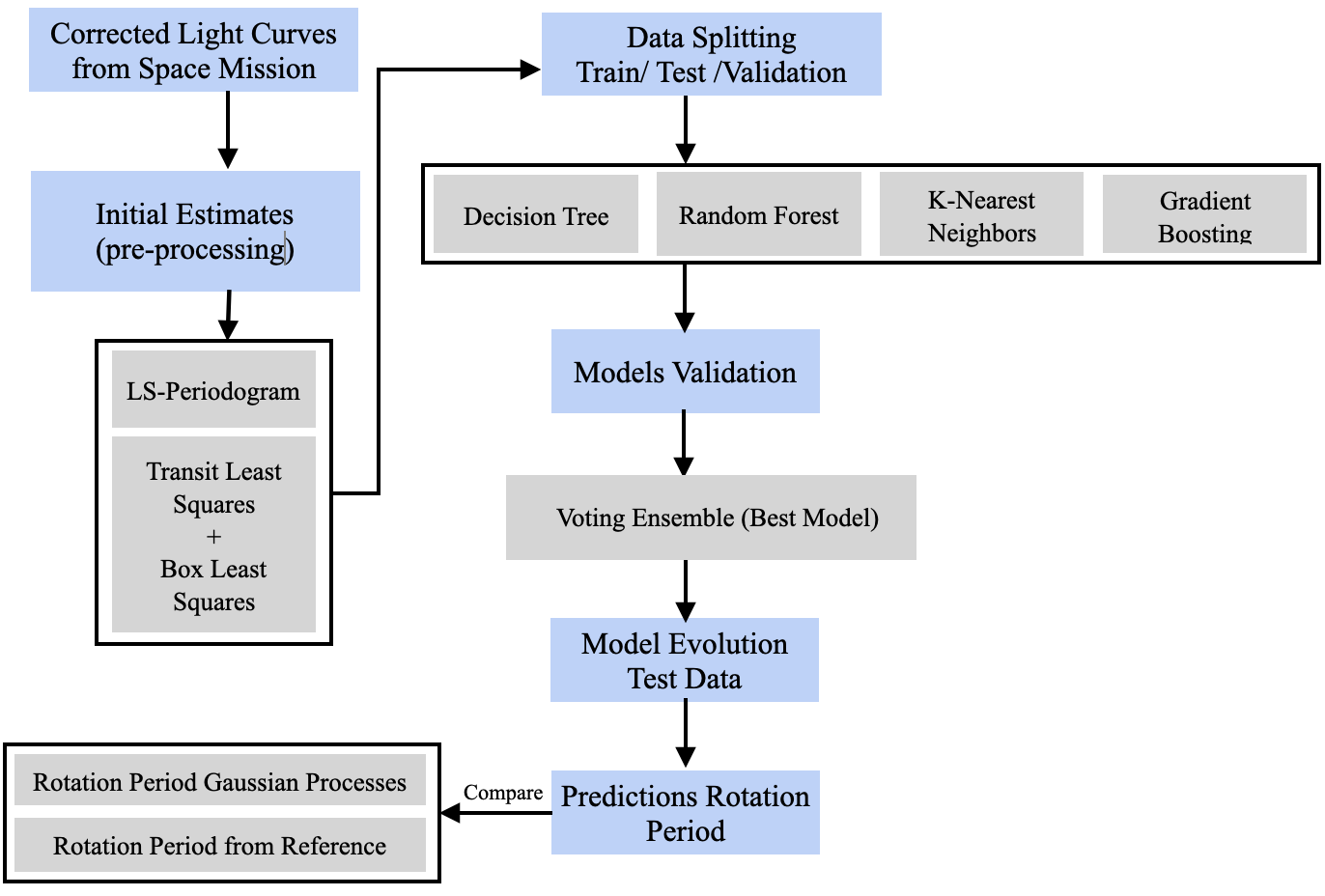}
    \caption{The diagram outlines the different steps involved, including initial estimates from corrected light curves, data splitting, application of machine learning models such as Decision Tree, Random Forest, K-Nearest Neighbors, and Gradient Boosting, model validation, testing, and the use of a Voting Ensemble (Best Model) approach. It also includes comparisons to the Gaussian Process approach for estimating rotation periods.
      }
         \label{fig:pipline}
   \end{figure*}

\begin{figure*}
\centering
\includegraphics[width=0.7\linewidth]{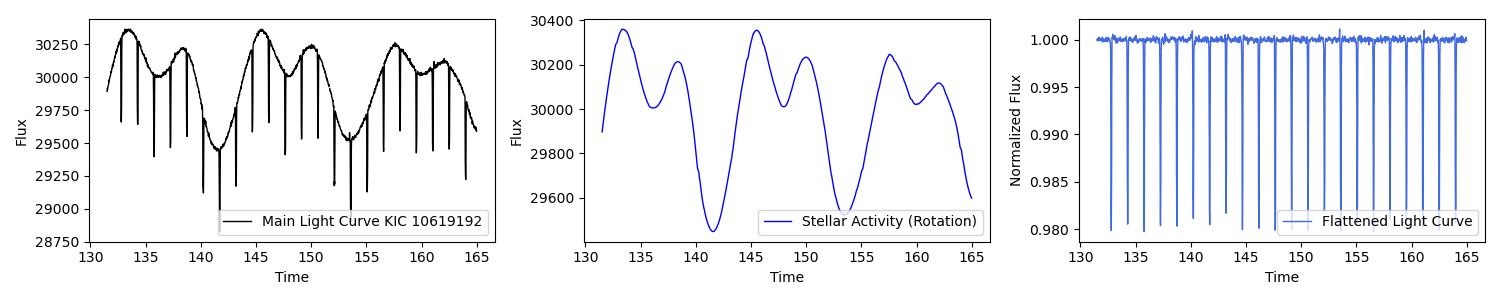}
\includegraphics[width=0.7\linewidth]{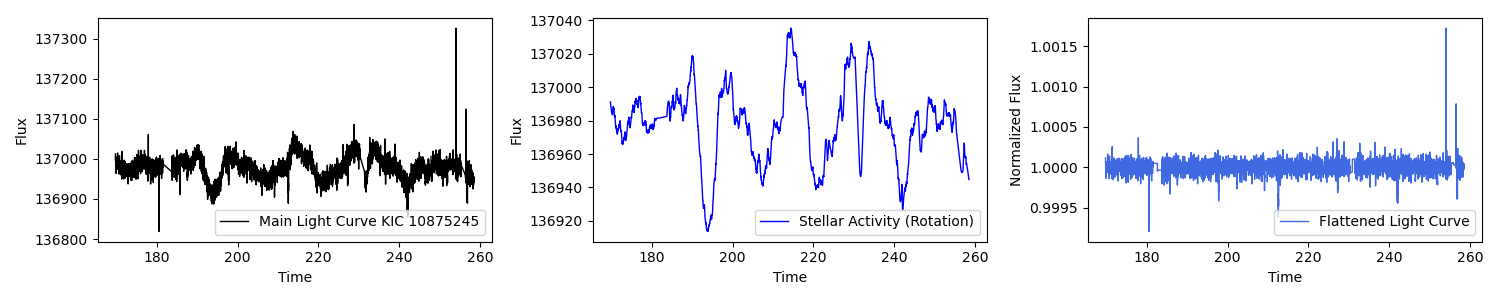}
\includegraphics[width=0.7\linewidth]{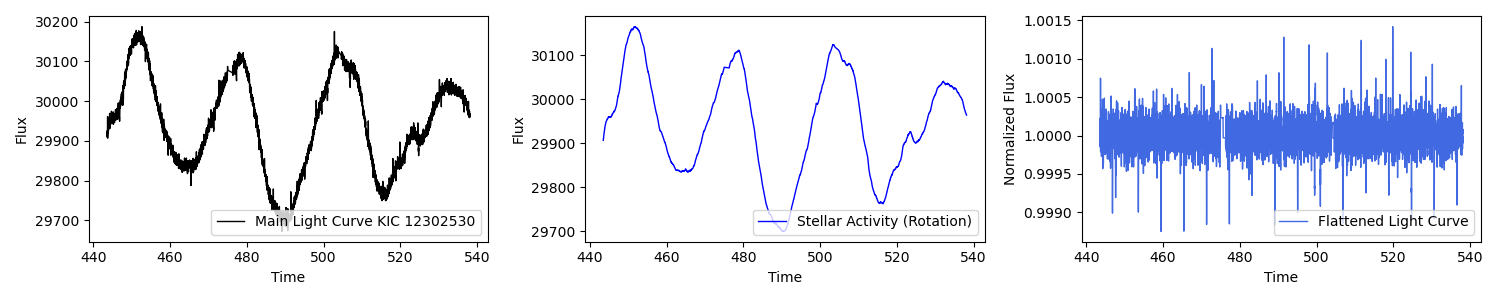}
\includegraphics[width=0.7\linewidth]{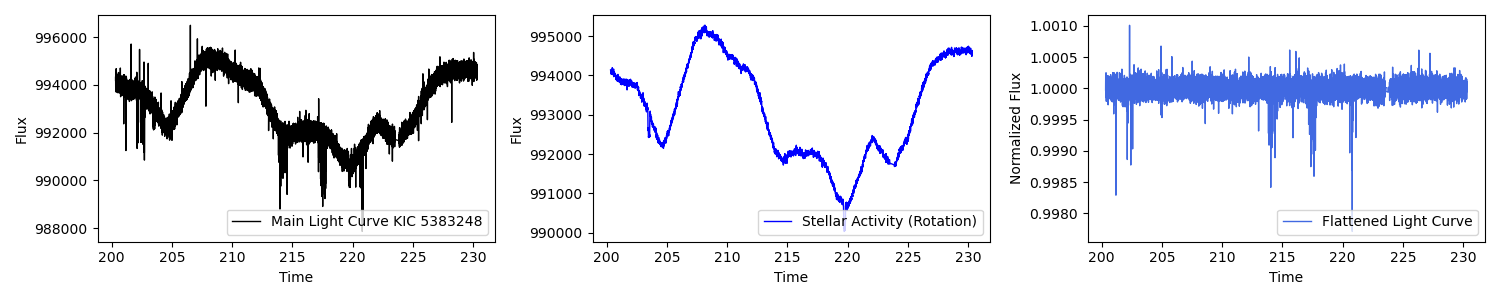}
\includegraphics[width=0.7\linewidth]{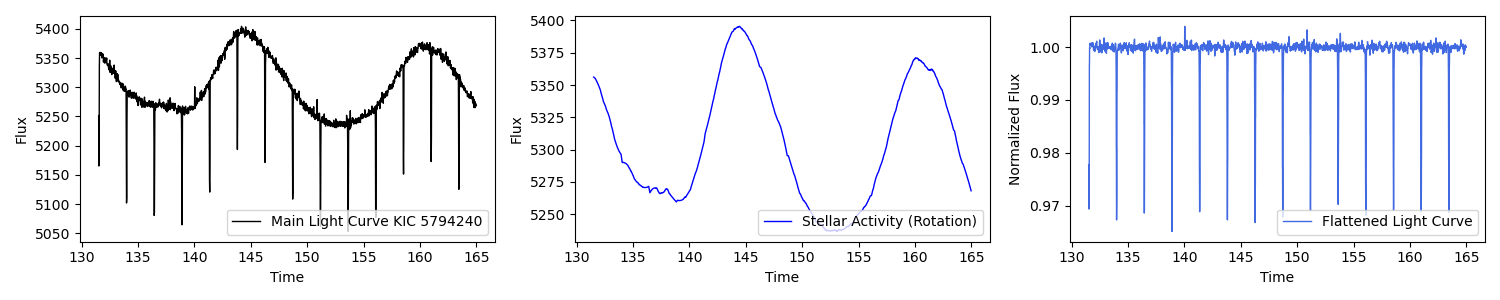}
\includegraphics[width=0.7\linewidth]{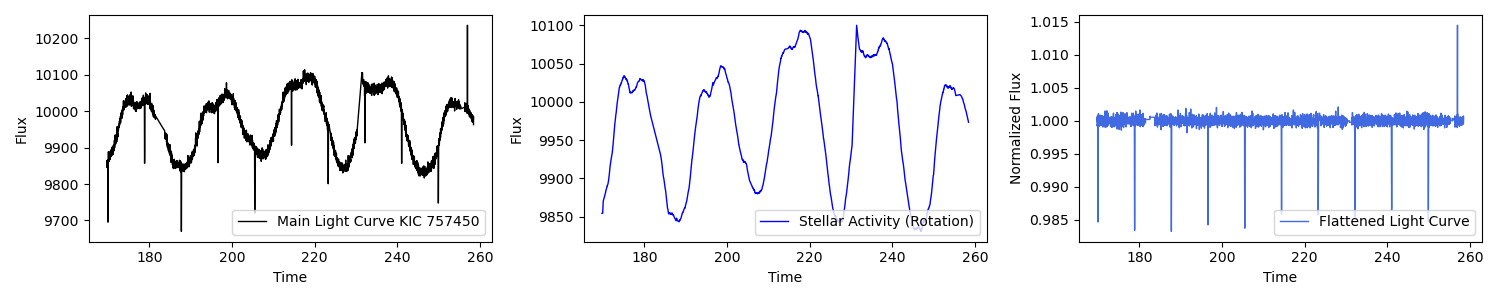}
\includegraphics[width=0.7\linewidth]{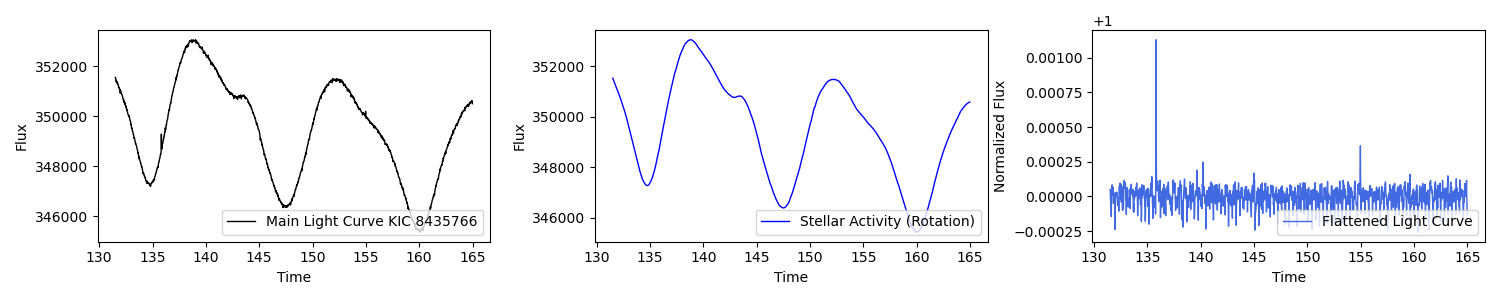}
\includegraphics[width=0.7\linewidth]{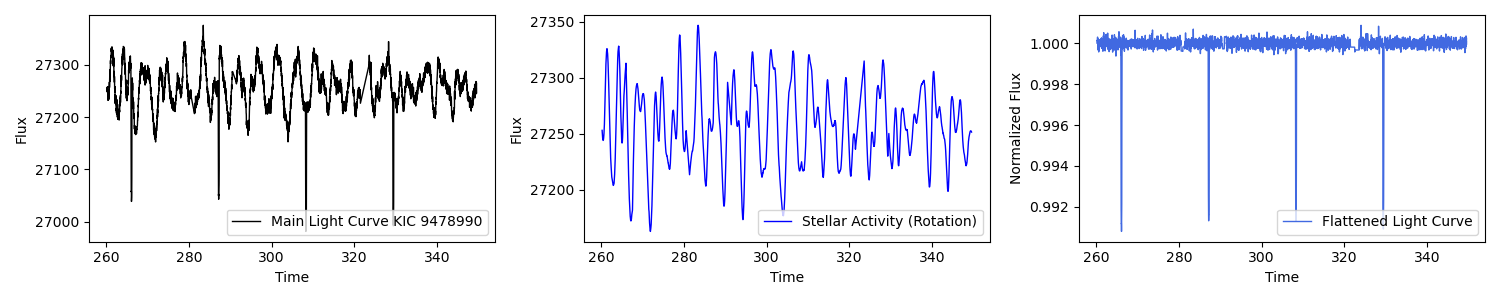}
\includegraphics[width=0.7\linewidth]{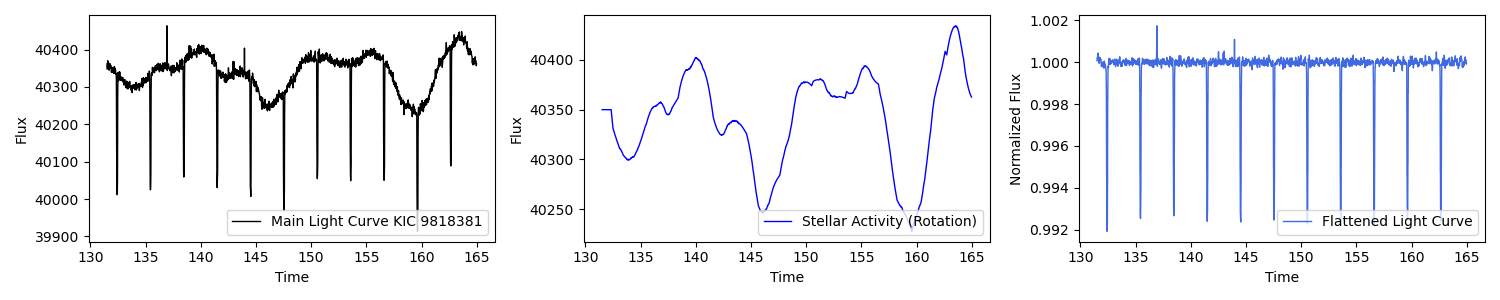}
\caption{The light curve of star which is divided into three panels. The first panel shows the raw light curve, which is a plot of the star's brightness as a function of time. The second panel shows the stellar activity rotation, which is a measure of the star's rotation rate. The third panel shows the flattened light curve, which is a plot of the star's brightness as a function of time after the stellar activity rotation has been removed.
              }
         \label{fig:stellar_flat}
   \end{figure*}
\section{Motivation}

Accurately measuring the rotation periods of exoplanet host stars is essential for understanding these planetary systems. Traditional periodogram techniques often struggle to effectively handle the noise, instrumental artifacts, and stellar variability present in light curve data. This study aims to address these challenges by employing advanced machine learning models, including Decision Tree, Random Forest, k-nearest Neighbors, and Gradient Boosting, combined with a Voting Ensemble approach and essential data preprocessing steps. The goal is to improve signal clarity and prediction accuracy of stellar rotation periods.
Improved accuracy in measuring stellar rotation periods can have significant implications. Magnetic activity revealed through stellar rotation can offer valuable information about stellar magnetism \citep{Basri2010}. Additionally, stellar variability can provide information on the age, architecture, and history of planetary systems. Solar-like stars lose angular momentum over time due to braking by magnetized stellar winds, leading to well-established stellar-rotation-activity-age relations \citep{Skumanich1972}. This relationship forms the foundation of gyrochronology, where stellar age can be estimated using the color of the star and its rotation rate \citep{Barnes2007}. 

In general, this study aims to contribute to a better understanding of exoplanet systems and their underlying properties through improved accuracy of stellar rotation period measurements.

\section{Data} \label{sec:data}

The data for this study were obtained from the Kepler Space Telescope. This archive provides high-precision photometric light curves of stars, including various data products and corrected light curves essential for accurate analysis. Specifically, we utilized Pre-search Data Conditioning Simple Aperture Photometry (PDCSAP) light curves from the Kepler Data Release 25 \citep[DR25,][]{kepler_dr25}, which offers the most comprehensive and corrected dataset available for stellar and exoplanetary research.

We selected a subset of Kepler IDs known for exhibiting clear exoplanet transits and stellar activity. The selection criteria included the quality of the light curves, the presence of well-defined transit events, and the availability of supplementary stellar parameters. The selected dataset includes light curves for the following Kepler IDs: KIC 10875245, KIC 12302530, KIC 10619192, KIC 9478990, KIC 9818381, KIC 5794240, KIC 757450, KIC 8435766, and KIC 5383248 (see Table \ref{tab:exoplanets}).
The raw light curves were preprocessed to remove instrumental artifacts and other sources of noise. This preprocessing involved several steps: normalization to remove long-term trends, detrending using a high-pass filter to eliminate stellar variability unrelated to transits, and outlier removal using a sigma-clipping algorithm to identify and remove outliers due to cosmic rays or instrumental glitches. We specifically used the PDCSAP \footnote{In the given passage, PDCSAP refers to Pre-search Data Conditioning Simple Aperture Photometry. It is a type of light curve data product provided by the Kepler Space Telescope mission. PDCSAP light curves have been processed to correct for common instrumental effects and systematic errors, such as spacecraft motion, thermal effects, and other instrumental artifacts. This correction process results in a cleaner dataset that is more suitable for detailed analysis of stellar and planetary signals. By using PDCSAP light curves, researchers can more accurately identify and characterize exoplanet transits and other astrophysical phenomena without the confounding influences of these systematic errors.} light curves, which are corrected for common instrumental effects and systematic errors, providing a cleaner dataset for analysis.

From the preprocessed light curves, we extracted key features relevant to exoplanet transit characterization, including transit parameters (duration, depth, and shape of the transit events) and stellar parameters (orbital periods, radii, and other relevant metrics for the host stars). To ensure robust model development and evaluation, the data were split into training, validation, and testing sets, with the training set consisting of \%70 of the data, the validation set \%15, and the testing set \%15. Data augmentation techniques, such as adding Gaussian noise and creating synthetic light curves, were applied to enhance the training dataset.

Additionally, we incorporated various data augmentation methods to further enhance the robustness of our dataset. By introducing synthetic variations in the transit parameters and adding Gaussian noise, we simulated a broader range of observational conditions and potential anomalies. This approach ensured that our machine learning (ML) models could generalize well to real-world data, accounting for potential noise and variability inherent in observational astronomy. This step was crucial in improving the models' resilience and accuracy when applied to actual Kepler light curves.
The comprehensive dataset, refined through meticulous preprocessing and augmentation, provided a solid foundation for developing and validating our machine-learning models. These models aimed to achieve precise period rotation prediction and stellar activity, significantly advancing the field of exoplanet research. Using the high-quality PDCSAP light curves and robust feature extraction methods, our study contributes to a deeper understanding of exoplanetary systems and their host stars, paving the way for future discoveries in the field.

Finally, we compare our results with a sample of stellar rotation periods determined by photometry and spectroscopy of the Kepler planet candidate host stars reported in \cite{Walkowicz2013}.

\begin{figure}
   \centering
   \includegraphics[width=1\linewidth]{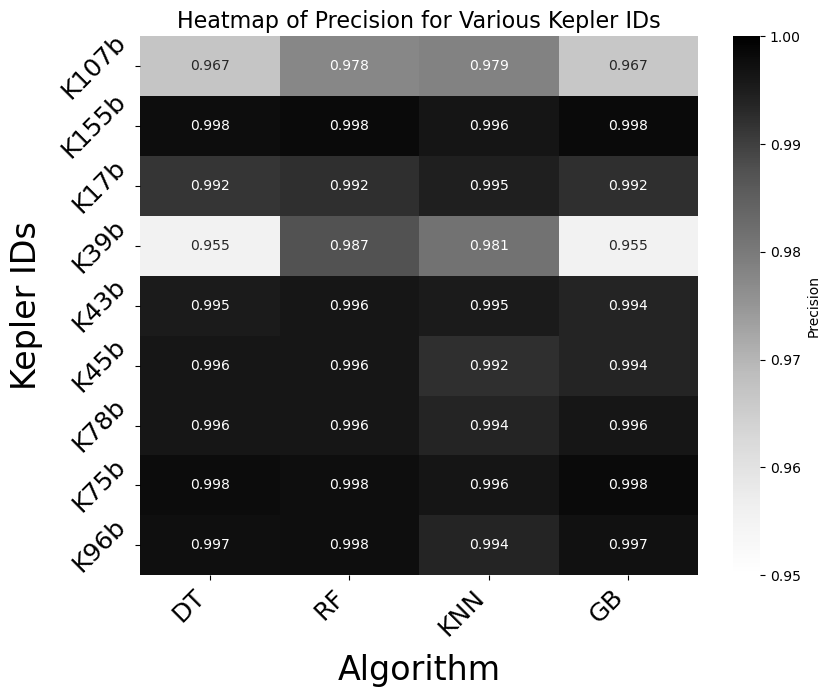}
   
    \caption{The heatmap displays the precision values achieved for each algorithm, with higher values represented by darker shades. The algorithms include Decision Tree (DT), Random Forest (RF), K-Nearest Neighbors (KNN), and Gradient Boosting (GB). Each Kepler ID (K107b, K155b, K17b, K39b, K43b, K45b, K78b, K75b, K96b) corresponds to a specific stellar activity. The values are rounded to three decimal places and indicate the precision of the respective algorithm on each Kepler ID.}

         \label{fig:Heat}
   \end{figure}
\begin{table*}[htbp]
\centering
\caption{This table shows the physical characteristics of various exoplanets orbiting Kepler stars. For each exoplanet, the table lists the Kepler ID, surface gravity, orbital period, radius (in Jupiter radii), and notes on the type of exoplanet. This detailed information is essential for comprehending the diversity and properties of exoplanetary systems observed by the Kepler Space Telescope}
\begin{tabular}{llcccc}
\hline
Exoplanet & Kepler ID & Surface Gravity (m/s²) & Period(day) & Radius(Jup) & Notes \\
\hline
Kepler-107 b & KIC 10875245 & 20.3 & 3.1800218 & 0.137 & Hot Jupiter exoplanet \\
Kepler-155 b & KIC 12302530 & 15.7 & 5.931194 & 0.186 & Rocky exoplanet \\
Kepler-17 b  & KIC 10619192 & 19.2 & 1.36 & 1.31 & Hot Jupiter exoplanet \\
Kepler-39 b  & KIC 9478990  & 3.4  & 1.4857108 & 1.24 &Super-Jupiter\\
Kepler-43 b  & KIC 9818381  & 4.1  & 3.0240949 & 1.219 & Hot Jupiter exoplanet \\
Kepler-45 b  & KIC 5794240  & 18.1 & 2.455239 & 0.96 &Hot Jupiter exoplanet \\
Kepler-75 b  & KIC 757450   & 30.0 & 8.8849116 & 1.05 &Super-Earth exoplanet \\
Kepler-78 b  & KIC 8435766  & 13.8 & 0.35500744 & 0.1 &Rocky exoplanet \\
Kepler-96 b  & KIC 5383248  & 5.6  & 16.2385 & 0.238 & Medium-sized short-period\\
\hline
\end{tabular}

\label{tab:exoplanets}
\end{table*}
\begin{figure}
   \centering
   \includegraphics[width=1\linewidth]{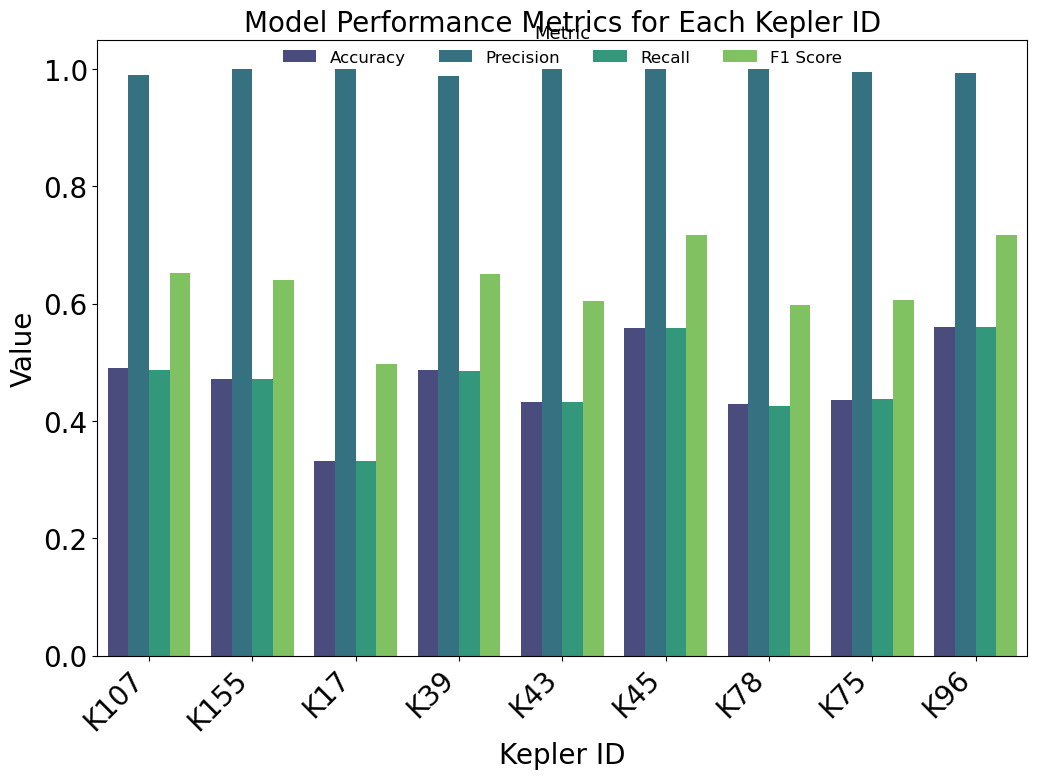}
      \caption{Bar plot illustrating the Best model performance metrics for each Kepler ID. The plot showcases the values of Accuracy, Precision, Recall, and F1 Score, which were calculated based on the model's confusion matrix. Each bar represents a specific Kepler ID, allowing for a visual comparison of the performance metrics across different IDs. The color-coded bars provide a clear distinction between the metrics, aiding in the assessment of the model's classification performance for each Kepler ID.
              }
         \label{fig:Barchart_Voting}
   \end{figure}

\begin{figure*}
\centering
\includegraphics[width=0.28\linewidth]{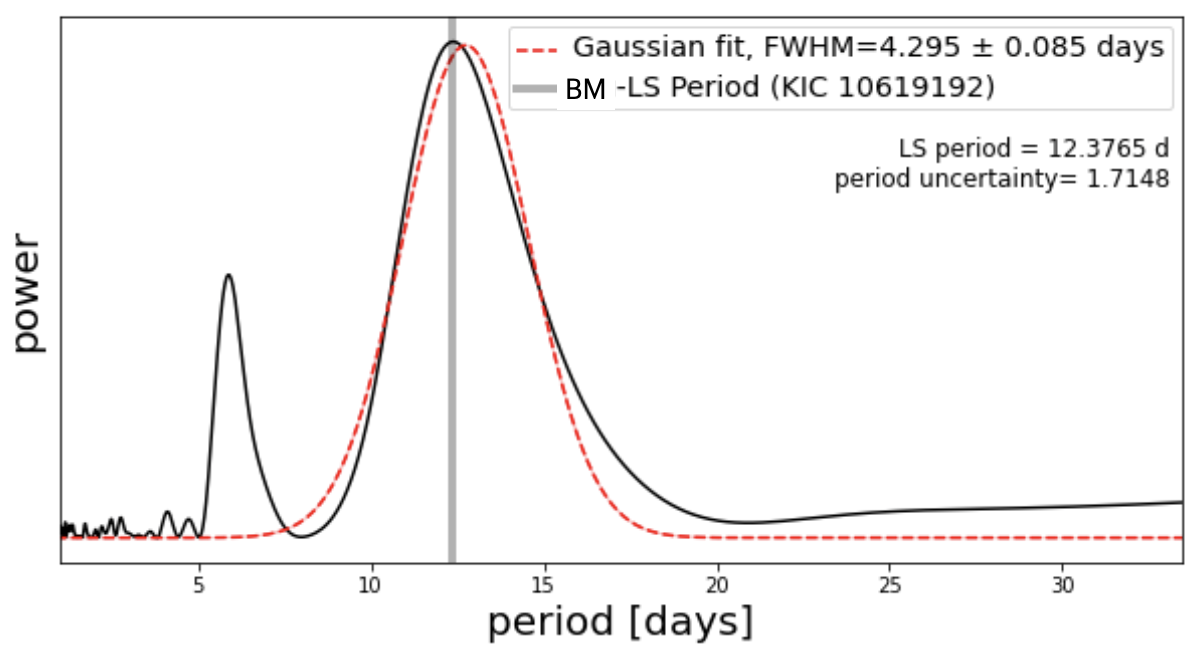}
\includegraphics[width=0.28\linewidth]{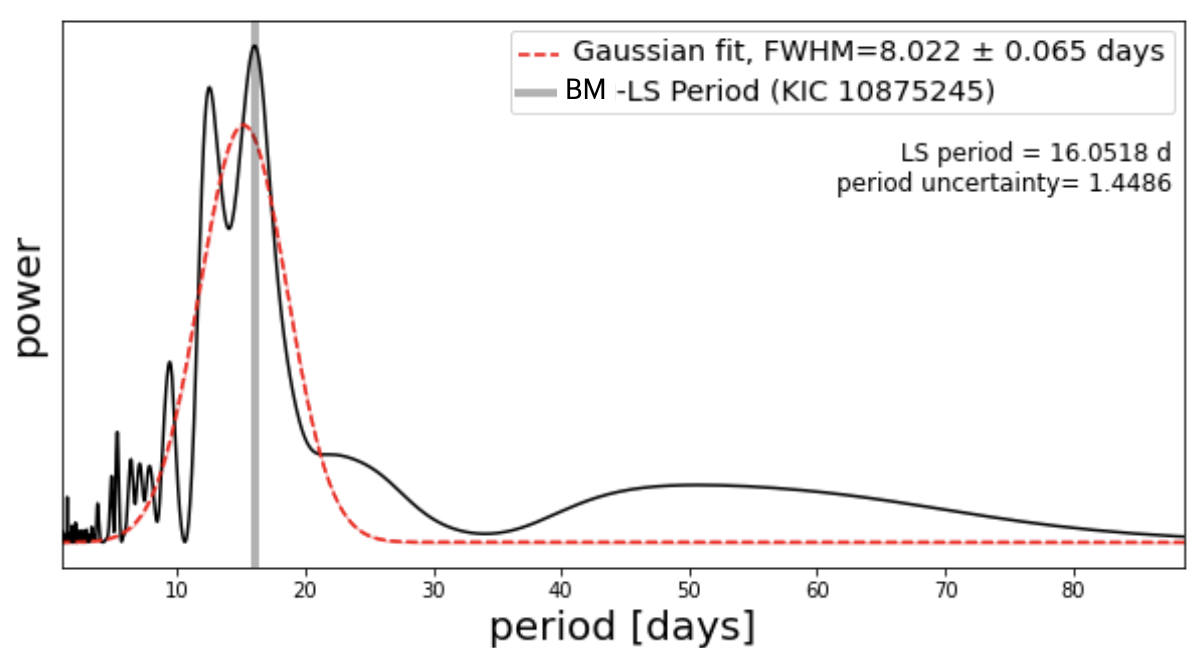}
\includegraphics[width=0.28\linewidth]{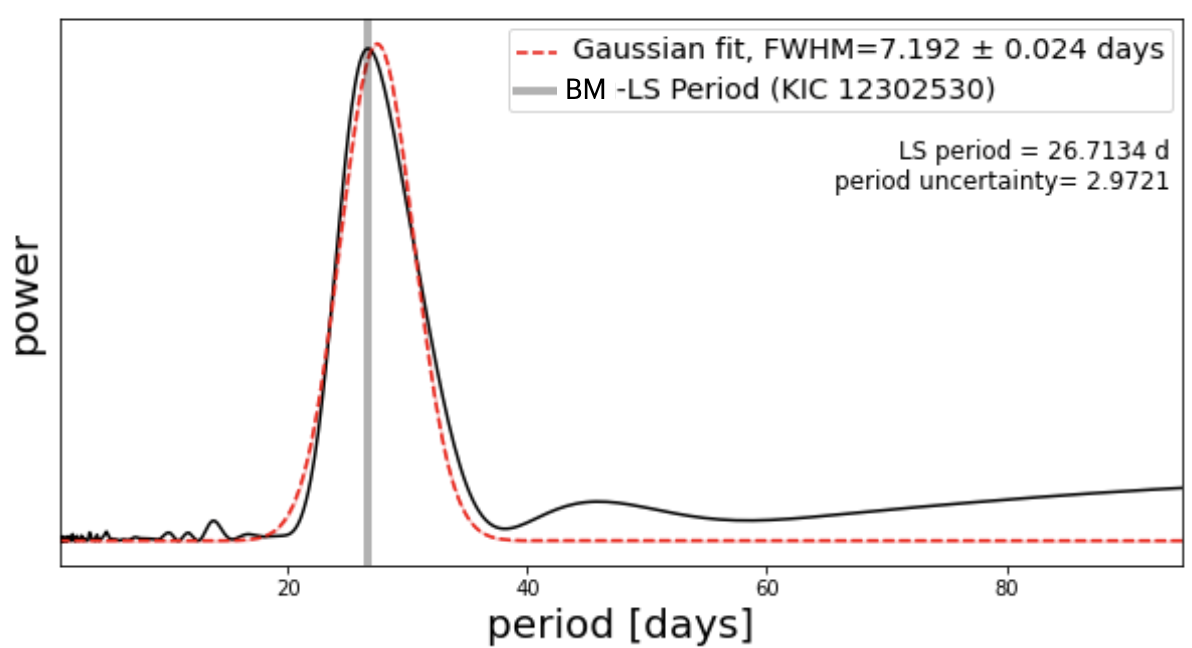}
\includegraphics[width=0.28\linewidth]{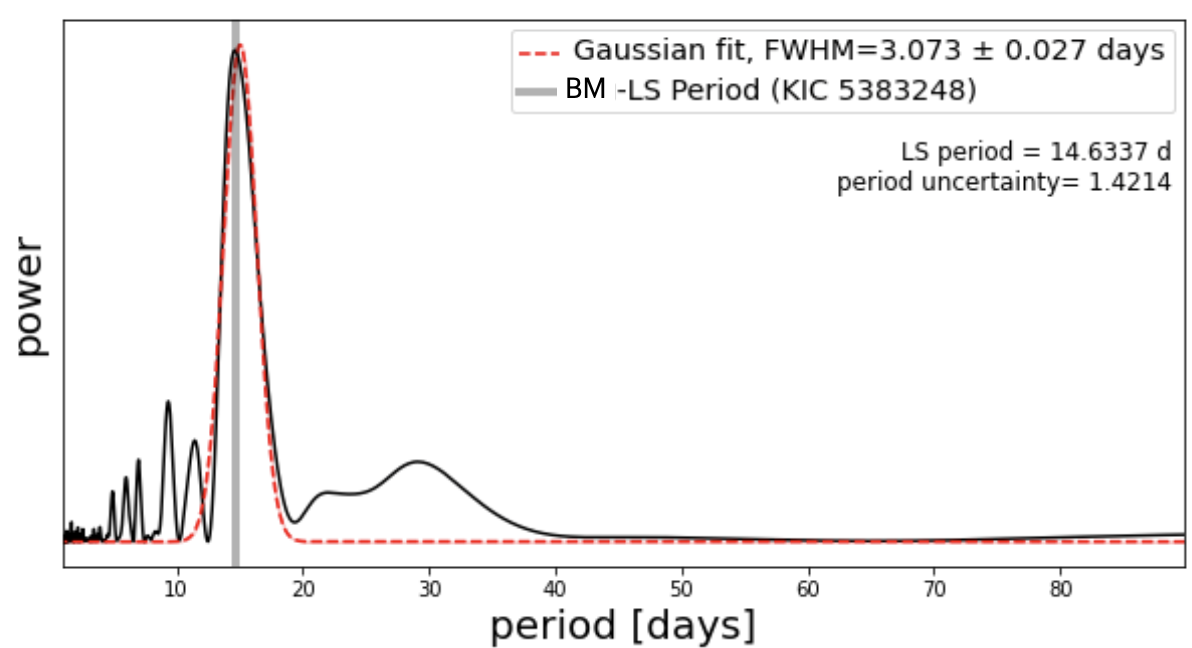}
\includegraphics[width=0.28\linewidth]{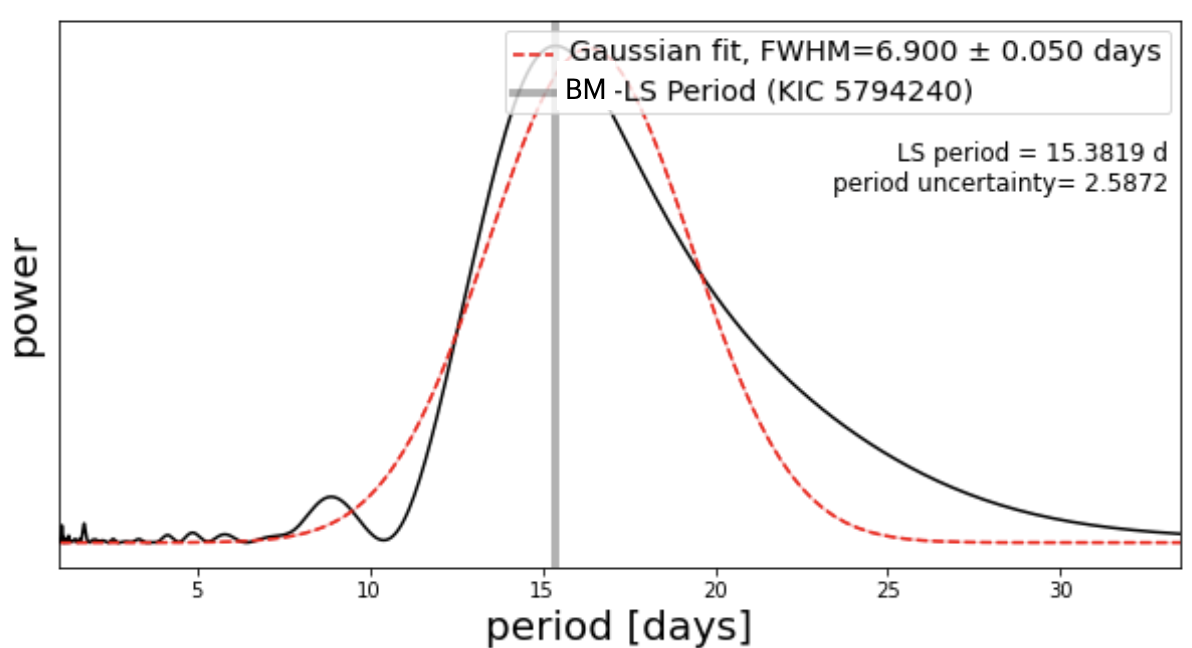}
\includegraphics[width=0.28\linewidth]{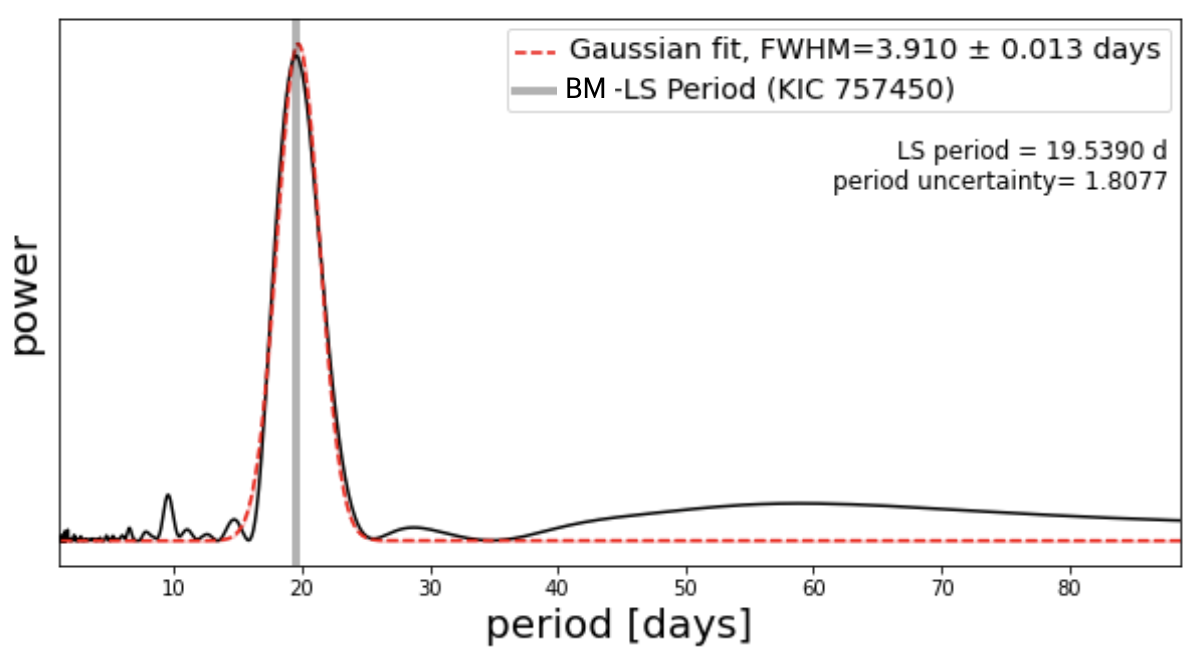}
\includegraphics[width=0.28\linewidth]{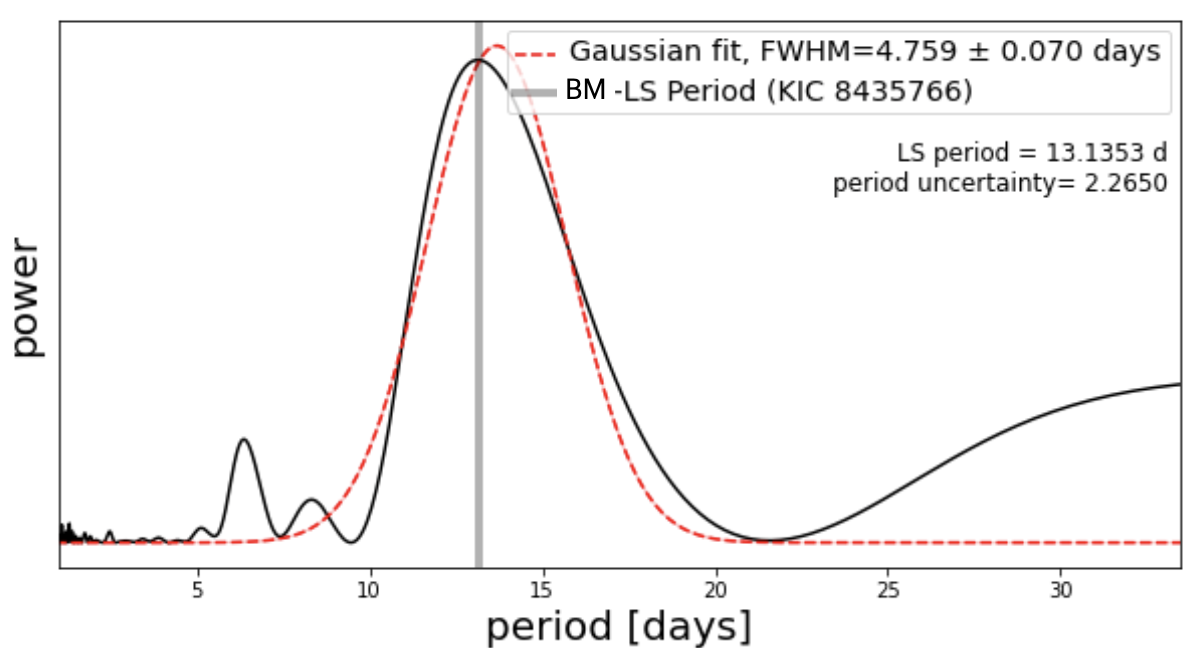}
\includegraphics[width=0.28\linewidth]{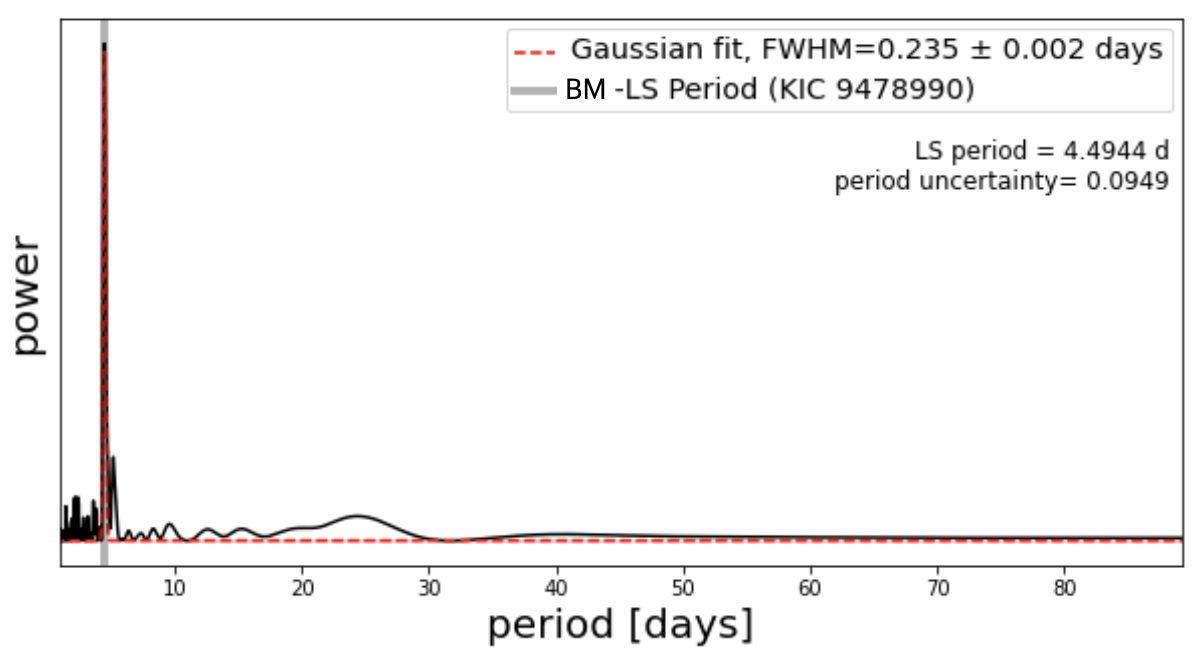}
\includegraphics[width=0.28\linewidth]{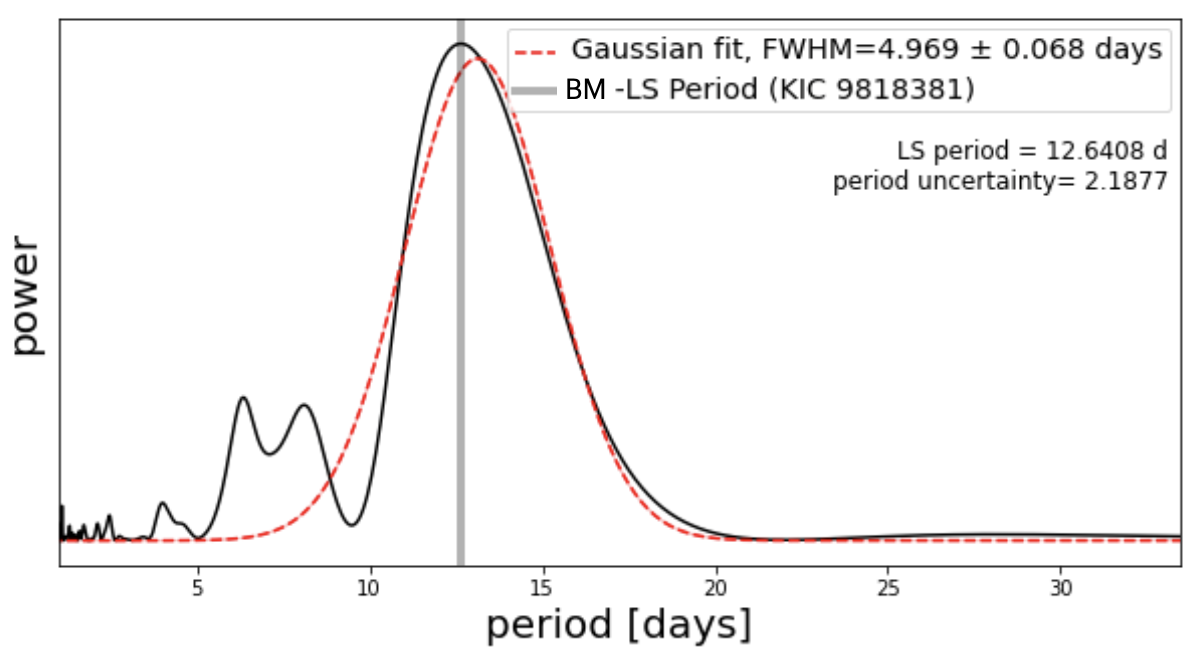}
\caption{The Lomb-Scargle periodograms computed from the photometric light curves of the exoplanet host stars in our dataset. The periodograms identify several statistically significant periodic signals present in the data, with the strongest peak (gray shaded line) in each plot corresponding to the estimated stellar rotation period. To quantify the uncertainty in these period estimates, we have fitted a Gaussian function to the primary peak and determined the half-width at half-maximum (HWHM) of the Gaussian (red-dashed line). This HWHM value provides an estimate of the period uncertainty that accounts for the sampling characteristics of the periodogram.
              }
         \label{fig:power_period}
   \end{figure*}
\section{Method}
\subsection{Data Preprocessing}

The raw light curves obtained from the Kepler Space Telescope required extensive preprocessing to ensure accurate analysis and modeling. This preprocessing involved several critical steps: normalization, detrending, outlier removal, and extraction of stellar rotation periods. Specifically, we utilized Pre-search Data Conditioning Simple Aperture Photometry (PDCSAP) light curves from the Kepler Data Release 25 (DR25).

Normalization was performed to remove long-term trends and bring consistency across different observations. Each light curve \( L(t) \) was normalized by subtracting the mean \( \mu \) and dividing by the standard deviation \( \sigma \), as shown in the following equation:

\begin{equation}
L_{\text{norm}}(t) = \frac{L(t) - \mu}{\sigma}
\label{eq:normalization}
\end{equation}

Detrending was conducted to eliminate stellar variability unrelated to transits. A high-pass filter was applied to the normalized light curves to remove these long-term trends. The high-pass filter can be represented by:

\begin{equation}
L_{\text{detrended}}(t) = L_{\text{norm}}(t) - L_{\text{trend}}(t)
\label{eq:detrending}
\end{equation}

where \( L_{\text{trend}}(t) \) is the low-frequency component of the light curve, typically extracted using a moving average or a polynomial fitting method.
Outliers due to cosmic rays or instrumental glitches were identified and removed using a sigma-clipping algorithm. The sigma-clipping method involves removing data points that deviate more than \( n \) standard deviations (\( n \sigma \)) from the median value of the light curve segment. The procedure can be mathematically expressed as:

\begin{equation}
L_{\text{clean}}(t) = \{ L_{\text{detrended}}(t) \mid |L_{\text{detrended}}(t) -\text{median}(L_{\text{detrended}})| < n \sigma \}
\label{eq:sigma_clipping}
\end{equation}

Accurate determination of stellar rotation periods is essential for distinguishing stellar activity from transit signals. We employed the Lomb-Scargle periodogram to estimate the rotation periods from the cleaned light curves. The Lomb-Scargle periodogram is suitable for unevenly spaced data and is given by:


\begin{equation}
\resizebox{0.45\textwidth}{!}{%
$P(\omega) = \frac{1}{2} \left[ \frac{\left[ \sum_{i} (L_{\text{clean}}(t_i) - \bar{L}) \cos \omega (t_i - \tau) \right]^2}{\sum_{i} \cos^2 \omega (t_i - \tau)} + \frac{\left[ \sum_{i} (L_{\text{clean}}(t_i) - \bar{L}) \sin \omega (t_i - \tau) \right]^2}{\sum_{i} \sin^2 \omega (t_i - \tau)} \right]$
}
\label{eq:lomb_scargle}
\end{equation}

where \( \omega \) is the angular frequency, \( t_i \) are the time points, \( \bar{L} \) is the mean of the light curve, and \( \tau \) is defined by:

where \( \omega \) is the angular frequency, \( t_i \) are the time points, \( \bar{L} \) is the mean of the light curve, and \( \tau \) is defined by:

\begin{equation}
\tan(2\omega \tau) = \frac{\sum_{i} \sin 2\omega t_i}{\sum_{i} \cos 2\omega t_i}
\label{eq:tau}
\end{equation}

The period corresponding to the highest peak in the periodogram \( P(\omega) \) was selected as the estimated rotation period of the star. After completing the preprocessing steps, we obtained clean and normalized light curves with extracted features, including transit parameters (duration, depth, and shape) and stellar parameters (orbital periods, radii, and rotation periods). These preprocessed light curves formed the basis for our subsequent ML model development and validation. By ensuring high-quality data through rigorous preprocessing, we aimed to improve the accuracy and robustness of our ML models, ultimately enhancing the period rotation prediction of exoplanet transits and stellar activity \footnote{https://github.com/FFazel13/ExoRotML.git}.

\subsection{Machine Learning models}

Having carefully prepared the data, our focus now shifts to the core of our analysis – the utilization of ML models for characterizing exoplanet transits in Kepler light curves. The selection of appropriate models is paramount to achieving accurate results in period rotation prediction. To inform our decision, we draw upon the insights gained from a previous study \citep{Fazel2023}. This study extensively explored the performance of different ML models on tasks akin to ours. Building upon their findings, we have chosen four highly promising models that have demonstrated a consistent record of success, as described below:
We employed several ML models to characterize exoplanet transits and determine stellar rotation periods from the preprocessed Kepler light curves. The models used include a Decision Tree (DT), Random Forest (RF), k-nearest Neighbors (KNN), Gradient Boosting (GB), and a Best-Model(BM) approach that combines multiple models.

\subsection{Gaussian processing model} \label{sec:DT}

In the past decade, Gaussian Processes (GPs) have become popular models for modeling both instrumental and astrophysical noise in light curves to improve the period rotation prediction. In this work, we apply scalable GP models using the software celerite to characterize exoplanet transits and stellar activity in Kepler lightcurves \citep{Foreman-Mackey2017}. We aim to build GP models that can retrieve accurate transit and rotation parameters, with a focus on the radius of the planet ($R_p$) and the rotation period of the star ($P_{\text{rot}}$).

Furthermore, we investigate the extent to which joint modeling improves overall period rotation prediction. We can express the GP model as:

\begin{equation}
f(t) \sim \text{GP}(\mu(t), k(t, t')), \label{eq:gp}
\end{equation}
where $\mu(t)$ is the mean function and $k(t, t')$ is the kernel function that captures the noise and stellar activity.

We develop a pipeline that automates multiple stages of data analysis. First, the pipeline preprocesses the lightcurves, which involves selecting PDCSAP instrumental correction \citep{LightkurveCollaboration2018}. Second, the pipeline makes initial estimates of the parameters using traditional Physics techniques. We use the Transit Least Squares algorithm by \citep{Hippke2019} to detect the transits and estimate the planet radius ($R_p$), transit period ($P_{\text{transit}}$), and the first transit time ($t_0$), and the Lomb-Scargle Periodogram by to estimate the stellar rotation period ($P_{\text{rot}}$). Third, the pipeline builds the GP model using the exoplanet toolkit so that the mean function captures the transit signals and the kernel function captures the noise and stellar activity \citep{Foreman-Mackey2020}. We use starry's limb-darkened lightcurve to model the transit signals, celerite's stochastically-driven damped harmonic oscillator (SHO) kernel to model background noise, and celerite's quasi-periodic Rotation kernel to model rotational modulation \citep{Luger2019}. The GP model is trained on the lightcurve data, providing a posterior distribution over the parameters. Fourth, the pipeline performs MCMC sampling using PyMC3's No U-Turn (NUTS) sampler to approximate the posterior distributions of the parameters \citep{Salvatier2016}.

\subsection{Decision Tree} \label{sec:DT}
A Decision Tree (DT) is a popular non-parametric supervised learning algorithm for regression and classification tasks. It models the relationships between features and the target variable by making a sequence of decisions based on the feature values \citep{Furnkranz2010}.
Starting from the root node, the algorithm recursively partitions the data based on the values of different features. At each node, the algorithm selects the best feature and the corresponding splitting criterion to maximize the homogeneity of the target variable within each partition. The commonly used splitting criteria include information gain, Gini impurity, and variance reduction.

Let $X$ represent the input features and $y$ represent the target variable. The decision tree algorithm seeks to find the best binary splitting rule at each internal node to minimize the impurity measure $I$. The impurity measure can be defined as:

\begin{equation}
    I(X_m) = \sum_{c \in C} p(c|X_m) \cdot (1 - p(c|X_m)),
    \label{eq:impurity}
\end{equation}
where $X_m$ is the subset of data at node $m$, $C$ represents the set of classes, and $p(c|X_m)$ is the proportion of class $c$ in node $m$.
The tree is built by repeatedly splitting the data based on the selected features until a stopping criterion is met. This criterion can be a predefined maximum depth of the tree, a minimum number of instances required to split a node, or a threshold on the impurity measure. Once the stopping criterion is reached, each leaf node is assigned a predicted value for regression tasks or a class label for classification tasks \citep{Hastie2010}.
Decision Trees offer several advantages, such as interpretability, handling both numerical and categorical features, and capturing non-linear relationships. However, they are prone to overfitting when the tree becomes too complex and may not generalize well to unseen data. To address this, techniques like pruning, setting the maximum depth, and using ensemble methods like Random Forests can be employed.

The Decision Tree algorithm has hyperparameters that can be tuned to optimize performance. The important hyperparameters include the maximum depth of the tree, the criterion used for splitting, the minimum number of samples required to split an internal node, and the minimum number of samples required to be at a leaf node. For hyperparameter tuning of a Decision Tree, key parameters include maximum tree depth, splitting criterion (e.g., Gini impurity, entropy), minimum samples required to split an internal node, and minimum samples required at a leaf node. Using grid search, for example, the optimal configuration for period prediction might be a maximum depth of 10, Gini criterion, and a minimum of 4 samples per split and leaf, which balances complexity and prevents overfitting.

Decision trees can be used as a powerful tool for period prediction by extracting relevant features from the observed light curves and making predictions about the characteristics of exoplanets. \citep{Buchhave2012}

\subsection{Random Forest} \label{sec:RF}
Random Forest (RF) is an ensemble learning method that combines multiple decision trees to create a robust and accurate predictive model \citep{Breiman2001}. It is widely used for both regression and classification tasks. Random Forest can be summarized as follows:

First, a user-defined number of decision trees, denoted as $N$, is created. For each tree $i$ in the ensemble:

\begin{enumerate}
    \item Randomly select a subset of the original training data with replacement, known as a bootstrap sample. This results in a dataset of the same size as the original, but with some duplicate and some omitted instances.
    \item Randomly select a subset of the features from the original feature set. This provides a diverse set of features for each tree and helps reduce correlation among the trees.
    \item Construct a decision tree using the bootstrap sample and the selected features. The tree is grown by recursively partitioning the data based on the selected features, optimizing a given criterion (e.g., Gini index for classification or mean squared error for regression) at each split.
\end{enumerate}

The prediction of a Random Forest for regression tasks is obtained by averaging the predictions of all individual trees:

\begin{equation}
    \hat{y} = \frac{1}{N} \sum_{i=1}^{N} f_i(x),
    \label{eq:rf_regression}
\end{equation}
where $\hat{y}$ is the predicted value, $N$ is the number of trees, and $f_i(x)$ is the prediction of the $i$-th tree for input $x$.\citep{Breiman2001}

For classification tasks, Random Forests use either the majority vote or the class probabilities from all trees to make the final prediction. The class probabilities can be estimated as:

\begin{equation}
    P(y=c | x) = \frac{1}{N} \sum_{i=1}^{N} P(y=c | x, T_i),
    \label{eq:rf_classification}
\end{equation}
where $P(y=c | x)$ is the probability of class $c$ given input $x$, $N$ is the number of trees, and $P(y=c | x, T_i)$ is the probability of class $c$ given input $x$ in the $i$-th tree.\citep{Breiman2001}

Random Forests are known for their ability to handle high-dimensional datasets, handle missing values, and provide estimates of feature importance. These properties make them popular in various domains, including bioinformatics, finance, and remote sensing.

The Random Forest algorithm has important hyperparameters that can be optimized for better performance, including the number of trees, maximum tree depth, number of features considered at each split, and the criterion for evaluating splits. Hyperparameter tuning techniques like grid search or random search can be used to find the best combination of hyperparameters based on performance on a validation set \citep{Pedregosa2011}.

Random Forest is a popular choice for exoplanet period rotation due to its robustness, feature extraction capabilities, ability to model non-linear relationships, generalization to unseen data, and interpretability. In Random Forests, important hyperparameters are the number of trees, maximum depth, number of features considered at each split, and the criterion for split evaluation. Grid search might find that the best setup for exoplanet transit characterization includes 100 trees, a maximum depth of 20, the square root of features per split, and the Gini criterion, enhancing model generalization and reducing overfitting. By combining multiple decision trees, Random Forest can handle noisy data, extract relevant features from light curves, capture complex relationships, generalize well, and provide insights into the factors influencing transit characteristics. These advantages make Random Forest a valuable tool for accurately characterizing exoplanet transits \citep{Buchhave2012}.

\subsection{Gradient Boosting} \label{sec:GB}
Gradient Boosting is an ensemble learning method that combines multiple weak learners to create a strong predictive model. It is particularly effective in solving regression and classification problems \citep{Friedman2001}. The Gradient Boosting algorithm can be summarized as follows: Initialize the model with a constant value, typically the mean of the target variable. For each iteration $m=1$ to $M$, compute the negative gradient of the loss function with respect to the current model's predictions:

\begin{equation}
    r_{im} = -\left[\frac{\partial L(y_i, F(x_i))}{\partial F(x_i)}\right]_{F(x) = F_{m-1}(x)}, 
    \label{eq:gradient_boosting1}
\end{equation}
where $y_i$ is the target variable, $F(x_i)$ is the current model's prediction for sample $x_i$, and $L(\cdot)$ is the loss function. Fit a weak learner, such as a decision tree, to the negative gradient values $r_{im}$, and obtain the corresponding prediction function $h(x)$. Update the model by adding a scaled version of $h(x)$ to the current model's predictions:

\begin{equation}
    F_m(x) = F_{m-1}(x) + \eta \cdot h(x),
    \label{eq:gradient_boosting2}
\end{equation}
where $\eta$ is the learning rate, controlling the contribution of each weak learner. Finally, combine the predictions of all weak learners to finalize the model:

\begin{equation}
    F(x) = \sum_{m=1}^{M} \eta \cdot h(x).
    \label{eq:gradient_boosting3}
\end{equation}

The choice of loss function $L(\cdot)$ depends on the type of problem being solved. Common loss functions for regression problems include mean squared error (MSE) and mean absolute error (MAE), while for classification problems, cross-entropy loss is often used \citep{Friedman2001}.
Gradient Boosting is a powerful technique for period rotation that addresses the challenges of overfitting and generalization. Regularization techniques, such as shrinkage (learning rate) and tree-based regularization, are employed to mitigate overfitting by reducing the contribution of each weak learner and controlling the complexity of the model \citep{Friedman2001}. The performance of a Gradient Boosting model heavily relies on hyperparameter tuning, including the number of weak learners, learning rate, tree depth, and regularization parameters. Techniques like grid search, random search, and Bayesian optimization can be employed for effective hyperparameter tuning \citep{snoek2012}. For Gradient Boosting, tuning involves the number of boosting stages, learning rate, maximum tree depth, and the loss function. A grid search might determine that 500 stages, a learning rate of 0.01, a maximum depth of 4, and mean squared error as the loss function offer the best performance for period rotation, balancing model complexity and convergence speed. By combining multiple weak learners, Gradient Boosting can capture complex relationships, extract relevant features from light curves, and accurately predict transit characteristics. Its versatility in handling various types of transit data and accommodating different loss functions makes Gradient Boosting a popular and effective choice for period rotation \citep{Friedman2001}.

\subsection{K-Nearest Neighbors (KNN)}\label{sec:KNN}
K-Nearest Neighbors (KNN) is a simple yet powerful supervised learning algorithm used for both regression and classification tasks \citep{Cover1967}. It makes predictions based on the similarity of input data points to their k nearest neighbors. The KNN algorithm can be summarized as follows:

Given a training dataset consisting of input features $X$ and corresponding target values $y$, the KNN algorithm operates as follows:

For a new input $x_{\text{new}}$ that needs to be classified or predicted, the algorithm finds the k nearest neighbors in the training dataset based on a distance metric, commonly Euclidean distance. The distance between two data points $x_i$ and $x_j$ can be calculated as:

\begin{equation}
    d(x_i, x_j) = \sqrt{\sum_{k=1}^{n}(x_{ik} - x_{jk})^2},
    \label{eq:knn_distance}
\end{equation}
where $n$ is the number of features.
K-Nearest Neighbors (KNN) is a versatile algorithm used for both classification and regression tasks. For classification, it assigns the most frequent class label among the k nearest neighbors through majority voting, while for regression, it predicts the average of the target values of the k nearest neighbors \citep{Altman1992}. The choice of the hyperparameter k significantly impacts the algorithm's performance, with smaller values leading to more flexible but potentially noisy predictions, and larger values providing smoother but potentially biased predictions. The optimal value of k can be determined using techniques like cross-validation or grid search. KNN is a non-parametric algorithm that can handle complex decision boundaries and is effective even when the data is not linearly separable. However, it can be sensitive to the choice of distance metric and the presence of irrelevant or noisy features \citep{Hastie2010}. The computational complexity of KNN can be high, especially with larger training datasets, but this can be addressed using techniques such as kd-trees or ball trees to accelerate the nearest neighbor search.

K-Nearest Neighbors (KNN) is commonly used for period rotation due to its ability to handle irregular or sparse transit data and its simplicity. It can adapt well to irregularities in the data and capture local variations in light curves, making it effective for accurately characterizing exoplanet transits. KNN's non-parametric nature and lack of assumptions about data distribution allow it to handle complex decision boundaries and different types of transit signals. K-Nearest Neighbors (KNN) hyperparameters include the number of neighbors (k) and the distance metric (e.g., Euclidean). Optimal values found through grid search might be k=5 and the Euclidean distance metric for period rotation, balancing between bias and variance to achieve the best predictive accuracy. However, KNN's performance can be sensitive to the choice of distance metric and the presence of irrelevant or noisy features \citep{Hastie2010}. It also has a high computational complexity, which can be mitigated using techniques like kd-trees or ball trees to speed up the nearest neighbor search \citep{Friedman2001}.

\subsection{Voting-based Classifier Ensemble}\label{sec:VCE}
A voting-based classifier ensemble combines the predictions of multiple individual classifiers to make a final decision. Here, we describe a combination model that combines Gradient Boosting, Random Forest, Decision Tree, and k-nearest neighbors using a voting scheme. The ensemble can be summarized as follows:
Here $X$ represents the input features and $y$ represents the target variable. We have four base classifiers: Gradient Boosting (GB), Random Forest (RF), Decision Tree (DT), and k-nearest neighbors (KNN). Each base classifier $i$ provides a class prediction $y_i$ for a given input $x$. To combine the predictions of the base classifiers, we employ a majority voting scheme. The class with the highest number of votes is chosen as the final prediction. In case of a tie, a predefined tie-breaking strategy can be used.
For each input $x$, the majority voting ensemble prediction $y_{\text{ensemble}}$ is obtained as follows:

\begin{equation}
    y_{\text{ensemble}} = \arg\max_{c} \sum_{i} w_i \mathbb{I}(y_i = c),
    \label{eq:voting_ensemble}
\end{equation}
where $c$ represents the class labels, $w_i$ is the weight assigned to each base classifier $i$, and $\mathbb{I}(y_i = c)$ is the indicator function that evaluates to 1 if $y_i = c$ and 0 otherwise.
The weights $w_i$ can be assigned based on the performance of each base classifier. For example, weights can be determined using techniques such as accuracy scores on a validation set.
The combination of different classifiers in the ensemble, such as Gradient Boosting, Random Forest, Decision Tree, and k-nearest neighbors, provides diversity in terms of modeling approaches and underlying assumptions. This diversity can improve the ensemble's overall performance and robustness \citep{Kittler1998}.

It is worth noting that the specific choice of base classifiers and the voting scheme can be adapted based on the problem domain and the characteristics of the dataset. Additionally, hyperparameter tuning for each base classifier is essential to optimize individual classifier performance and, consequently, the performance of the ensemble.

  \begin{figure*}
   \centering
   \includegraphics[width=1\linewidth]{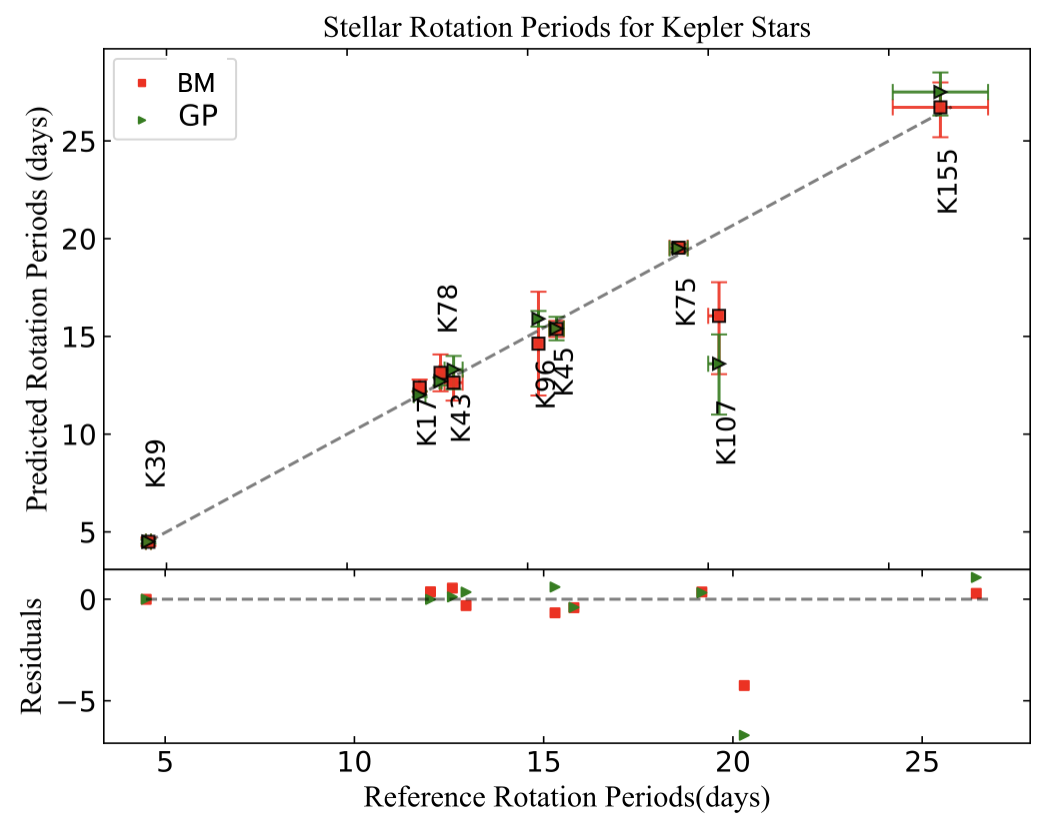}
      \caption{Comparison of predicted stellar rotation periods for Kepler stars using the Best Model (BM) and Gaussian Process model (GP). The top panel displays a scatter plot of predicted rotation periods against the reference rotation periods, with a 1:1 reference line (dashed) indicating perfect prediction. Each star is labeled by its Kepler ID. The bottom panel shows the residuals (difference between predicted and reference periods) for both models. The Best Model (BM) predictions are marked in red squares, and the Gaussian Process model (GP) predictions are marked in green triangles. The results indicate that the Best Model (BM) consistently provides more accurate rotation period estimates, as evidenced by the smaller residuals and closer alignment with the reference values. Error bars represent the uncertainties associated with each prediction.
              }
         \label{fig:preiodpredict}
   \end{figure*}
\subsection{Training Pipeline}

In Figure \ref{fig:pipline}, we demonstrate the different steps involved in estimating stellar rotation periods from light curves with exoplanet transits.
Using the Fourier transform algorithm, we characterize the trend and detrend the data to obtain the flux and time of the trend related to the stellar activity, which we then provide to machine learning models to get the best fit. Further, we use the fit corresponding to each model as the input for the Lomb-Scargle (LS) periodogram to obtain the maximum period and the error (FWHM/2), which we report as the exact period for each model.
The data is then split into training, validation, and testing sets. Various machine learning models are applied and validated, including Decision Tree, Random Forest, K-nearest neighbors, and Gradient Boosting. A Voting Ensemble approach is used to combine the best models for final predictions. The resulting stellar rotation periods are compared with those estimated by the Gaussian Process approach and reference values to ensure accuracy and reliability. This methodology aims to improve the precision of stellar rotation period estimations, which is essential for accurately characterizing exoplanet transits and distinguishing them from stellar activity.

The training pipeline for the ensemble is as follows (see Figure \ref{fig:pipline}). First, the labeled dataset is split into a training set and a validation set. Then, each base classifier is trained independently on the training set. This includes training the Gradient Boosting model (GB) using gradient boosting on decision trees, the Random Forest model (RF) using an ensemble of decision trees with random feature selection, the Decision Tree model (DT) using a single decision tree, and the k-nearest neighbors model (KNN) using the k-nearest neighbors algorithm.

Next, the performance of each base classifier is evaluated on the validation set, and weights are assigned to each classifier based on their performance. Techniques such as accuracy, or cross-validation scores on the validation set can be used to determine the weights. Higher-performing classifiers are assigned higher weights to have a greater influence in the voting scheme during the testing pipeline.

The testing pipeline for the ensemble is as follows. First, the class predictions from each base classifier are obtained for each test sample. Then, the voting scheme is applied to combine the predictions. The class with the highest number of votes is chosen as the final prediction for each test sample. The voting scheme can be a simple majority voting, where the class with the most votes is selected. In case of a tie, a predefined tie-breaking strategy can be used.

The ensemble benefits from the diversity of base classifiers, as each model captures different aspects of the data and may have varying strengths and weaknesses. By combining the predictions of best models, the ensemble can achieve improved overall performance and generalization. It is important to note that hyperparameter tuning for each base classifier is necessary to optimize individual classifier performance and, consequently, the performance of the ensemble. Techniques such as grid search or random search can be used to find the optimal hyperparameter settings for each model.

\begin{table*}
\centering
\scriptsize
\begin{tabular}{lcccccccccccc}
\hline
ID & rmse-DT & rmse-RF & rmse-KN & rmse-GB & rmse-BM & P-DT & P-RF & P-KN & P-GB & P-BM & P-GP & P-Ref \\ \hline
10875245 & 0.7361 & 0.5340 & 0.6448 & 4.6172 & 0.5430& 16.0 $\pm$ 1.5 & 16.0 $\pm$ 1.5 & 16.0 $\pm$ 1.4 & 16.1 $\pm$ 1.4 & 16.0 $\pm$ 1.4 & $13.6^{+2.6}_{-1.5}$ & $20^{+3.3}_{-0.01}$ \\
302530 & 0.7777 & 0.4859 & 0.5524 & 6.1664 &0.509 & 26.7 $\pm$ 3.0 & 26.7 $\pm$ 3.0 & 26.7 $\pm$ 3.0 & 26.7 $\pm$ 3.0 & 26.7 $\pm$ 3.0 & $27.5^{+1.2}_{-1}$ & $26.4^{+1.3}_{-1.3}$ \\ 
10619192 & 4.3479 & 2.6783 & 3.1398 & 13.2066 & 2.688 &  26.7 $\pm$ 1.7 & 26.7 $\pm$ 1.7 & 26.7 $\pm$ 1.7 & 26.7 $\pm$ 1.7 & 12.4 $\pm$ 1.7 & $12.0^{+0.1}_{-0.1}$ & $12.0^{+0.2}_{-0.2}$ \\ 
9478990 & 1.7536 & 1.2339 & 1.6824 & 16.7481 & 1.309& 4.5 $\pm$ 0.1 & 4.5 $\pm$ 0.1 & 4.5 $\pm$ 0.1 & 4.5 $\pm$ 0.1 & 4.5 $\pm$ 0.1 & $4.5^{+0.1}_{-0.1}$ & $4.5^{+0.07}_{-0.07}$ \\ 
9818381 & 0.8780 & 0.5846 & 0.5957 & 2.5127 & 0.558 & 12.6 $\pm$ 2.2 & 12.6 $\pm$ 2.2 & 12.6 $\pm$ 2.2 & 12.6 $\pm$ 2.2 & 12.6 $\pm$ 2.2 & $13.3^{+0.1}_{-0.7}$ & $13.0^{+0.2}_{-0.2}$ \\ 
5794240 & 0.5270 & 0.3455 & 0.4013 & 1.2045 & 0.346 & 15.4 $\pm$ 2.6 & 15.4 $\pm$ 2.6 & 15.4 $\pm$ 2.6 & 15.4 $\pm$ 2.6 & 15.4 $\pm$ 2.6 & $15.4^{+0.6}_{-0.6}$ & $15.8^{+0.2}_{-0.2}$ \\ 
8435766 & 24.6537 & 14.6678 & 17.6543 & 62.8880 & 14.338& 13.1 $\pm$ 2.3 & 13.1 $\pm$ 2.3 & 13.1 $\pm$ 2.3 & 13.1 $\pm$ 2.3 & 13.1 $\pm$ 2.3 & $12.7^{+0.2}_{-0.2}$ & $12.6^{+0.03}_{-0.03}$ \\ 
757450 & 0.6567 & 0.4249 & 0.5267 & 4.7397 & 0.427 & 19.5 $\pm$ 1.8 & 19.5 $\pm$ 1.8 & 19.5 $\pm$ 1.8 & 19.5 $\pm$ 1.8 & 19.5 $\pm$ 1.8 & $19.5^{+0.2}_{-0.2}$ & $19.2^{+0.2}_{-0.2}$ \\
5383248 & 16.0523 & 10.2492 & 12.2046 & 169.6816 & 10.66 & 14.6 $\pm$ 1.4 & 14.6 $\pm$ 1.4 & 14.6 $\pm$ 1.4 & 14.6 $\pm$ 1.4 & 14.6 $\pm$ 1.4 & $15.9^{+0.4}_{-0.4}$ & $15.3^{+0.01}_{-0.01}$ \\ 
\hline
\end{tabular}
\caption{The error bars reflect the model's internal estimate of prediction uncertainty and can be similar across different models due to consistent training data and estimation methods. RMSE, however, is a direct measure of prediction accuracy and varies between models depending on how well they capture the true patterns in the data. Thus, similar error bars do not imply similar RMSE values, highlighting the importance of considering both metrics to evaluate model performance comprehensively. Predicted stellar rotation periods and their associated root mean square errors (RMSE) for various models (Decision Tree, Random Forest, K-Nearest Neighbors, Gradient Boosting, and Best Model) are compared to reference periods (P-Ref) and Gaussian Process predictions (P-GP). Each model's predictions (P-DT, P-RF, P-KN, P-GB, P-BM) are displayed with uncertainties. This table highlights the performance of different models in predicting the rotation periods of Kepler stars.s}
\label{table:stellar_rotation}
\end{table*}

\begin{figure}
   \centering
   \includegraphics[width=1\linewidth]{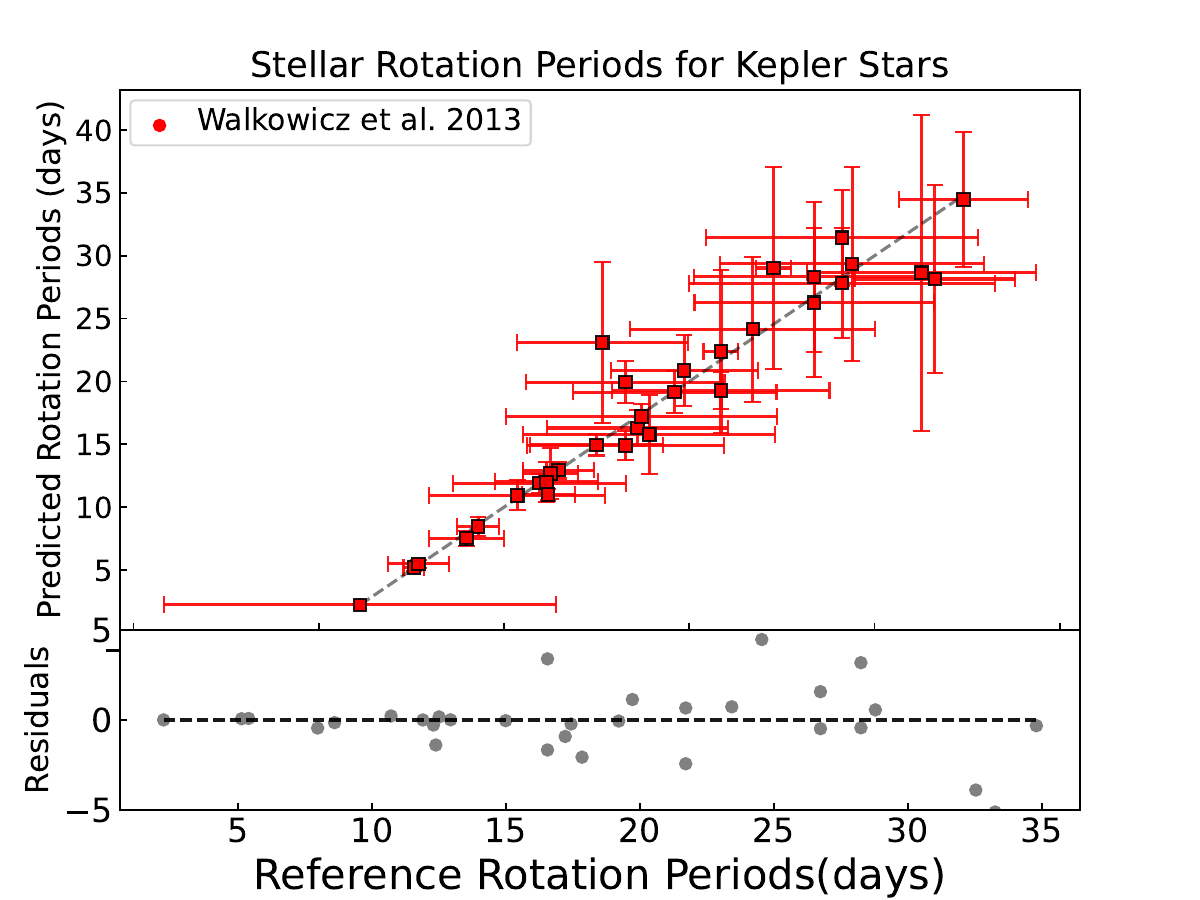}
      \caption{Predicted stellar rotation periods for Kepler stars using the Best Model (BM). The top panel displays the predicted rotation periods against the reference rotation periods, with a 1:1 line (dashed) indicating perfect prediction alignment. Red dots represent the predictions from \cite{Walkowicz2013}. 
      The bottom panel shows the residuals (differences between predicted and reference periods), with zero residual (dashed line) indicating perfect prediction. The results demonstrate the superior performance of the BM model, producing accurate rotation period estimates closely aligning with the reference values and exhibiting small residuals.
              }
         \label{fig:preiodpredictWalkowicz2013}
   \end{figure}

\section{Result}

In this study, the stellar rotation period, which is the time it takes for a star to complete one full rotation on its axis, was incorporated as a critical feature to enhance predictive models for exoplanet properties. Traditionally, the Gaussian Process Model (GPM) has been used to estimate stellar rotation from light curves. However, we introduced a novel approach using ML models trained on extensive datasets of known rotation periods. These models can more accurately identify dips in brightness caused by rotating star spots compared to GPM. This improved accuracy is essential for distinguishing between exoplanet transits and stellar activity. Our ML model, named Kepler Vision, can predict potential transit events based on its understanding of stellar rotation, allowing for more focused observations during the most promising times to detect a planet transiting its star. By integrating the stellar rotation period as a feature in the predictive models, we captured the rotational effects on transit characteristics, which improved the precision and reliability of exoplanet discoveries. 

\subsection{Exoplanet transits and Stellar activity}\label{sec:FW_stellar_activity}

We employ Fourier transform techniques to analyze the Kepler light curves and distinguish the signatures of exoplanet transits and stellar activity. The Fourier transform is a powerful tool that decomposes the time-domain light curve signal into its constituent frequency components.

The Fourier transform of the light curve \( f(t) \) is defined as:

\begin{equation}
\mathcal{F}(f)(w) = \int_{-\infty}^{\infty} f(t) e^{-i2\pi wt} dt,
\end{equation}
where \( w \) represents the angular frequency. The power spectrum, which provides the power of each frequency component, is calculated as:

\begin{equation}
P(w) = |\mathcal{F}(f)(w)|^2.
\end{equation}

Exoplanet transits appear in the light curve as periodic dips with characteristic transit duration and depth, corresponding to distinct frequency signatures in the power spectrum. The transit period \( P_{\text{transit}} \) is encoded as a peak at the frequency \( w_{\text{transit}} = \frac{2\pi}{P_{\text{transit}}} \). 
Additionally, the transit duration \( \tau_{\text{transit}} \) is inversely related to the width of this peak in the power spectrum.
Conversely, stellar activity such as star spots and stellar rotation manifest as quasi-periodic variations in the light curve. These signals are captured by broad, smooth peaks in the power spectrum, centered around the stellar rotation frequency \( w_{\text{rot}} = \frac{2\pi}{P_{\text{rot}}} \), where \( P_{\text{rot}} \) is the stellar rotation period.

As shown in Figure \ref{fig:stellar_flat}, by analyzing the power spectrum of the Kepler light curves, we can disentangle the contributions from exoplanet transits and stellar activity. This information can be used to refine the parameter estimates obtained from the Gaussian Process modeling, leading to a more comprehensive characterization of the exoplanet and its host star. This figure provides a clear illustration of the light curve analysis process, demonstrating how the Fourier transform helps distinguish between exoplanet transits and stellar activity.

The first column of Figure \ref{fig:stellar_flat} shows the original light curves for a selection of Kepler IDs (listed in table \ref{tab:exoplanets}), highlighting the raw flux measurements over time. These light curves exhibit a variety of features, including potential exoplanet transits and other stellar phenomena. The second column displays the stellar activity rotation signals isolated using Fourier transform techniques, which help in identifying the periodic variations due to starspots and stellar rotation. This step is necessary for differentiating between the intrinsic stellar activity and exoplanet transit signals. The third column presents the flattened light curves, where the rotational components have been removed, leaving behind a cleaner dataset that is more suitable for detecting and analyzing transit events.

Figure \ref{fig:stellar_flat} summarizes the effectiveness of our preprocessing methodology, which is a critical part of the data preprocessing stage in our pipeline (refer to the "Initial Estimates" and "Models Validation" stages in Figure \ref{fig:pipline}). By systematically addressing and removing the noise and stellar activity components, we have significantly enhanced the clarity of the transit signals. This improvement is essential for accurate period rotation properties and further analysis. The clear separation of raw light curves, stellar rotation signals, and flattened light curves demonstrates our approach's robustness and sets a solid foundation for applying advanced ML techniques to predict and validate stellar and exoplanet parameters.

\subsection{Light curve characterisations}
Beyond the Gaussian Process modeling approach, we also explore the use of other ML techniques to analyze the Kepler lightcurves and disentangle the signatures of exoplanet transits and stellar activity. We provide the best fit to the light curve using different ML algorithms such as Decision Tree, Random Forest, Gradient Boosting, K-Nearest Neighbors, and a Voting-based Classifier Ensemble. The respective root mean square error (RMSE) values for these models, as reported in Table \ref{table:stellar_rotation}, highlight their performance.

The Decision Tree and Random Forest models can be employed to classify the observed photometric variations in the Kepler lightcurves as either exoplanet transit or stellar activity. These models can learn the characteristic features of transit and activity signals, such as their periodic patterns, depths, and durations, and use this knowledge to make predictions on new, unseen light curves. The lower RMSE values for these models, compared to Gradient Boosting and K-Nearest Neighbors, suggest that they are better able to capture the relevant patterns in the data. For instance, the Decision Tree model shows an RMSE of 0.7361 for KIC 10875245 and similar low values for other KICs, indicating robust performance. In general, in BM model the RMSE was approximately 50\% lower than the DT model and performed comparably with RF model.

Gradient Boosting is another powerful ML technique that can be used to improve the period rotation prediction and stellar activity. By iteratively building an ensemble of weak predictive models (e.g., decision trees), Gradient Boosting can capture complex non-linear relationships in the data and provide more accurate parameter estimates compared to individual models. However, the relatively higher RMSE values, such as 4.6172 for KIC 10875245, indicate that it may struggle to fully disentangle the transit and activity signals in some cases.

The K-Nearest Neighbors (KNN) algorithm can be used to identify and classify the different patterns in the Kepler light curves corresponding to exoplanet transits and stellar activity. KNN relies on the similarity between the input lightcurve and a database of labeled examples, allowing it to make predictions without assuming any specific functional form for the underlying signals. The RMSE values, such as 2.67 for KIC 10619192, suggest that it can provide reasonably accurate period rotation but may not be as effective as the Decision Tree and Random Forest models. In general, in BM model the RMSE is approximately 17\% lower than the KNN model.

The results from the Gaussian Process modeling, as well as the various ML techniques explored, demonstrate the benefits of jointly modeling the exoplanet transits and stellar activity to obtain accurate parameter estimates. The joint GP model, in particular, has shown promising performance in recovering the stellar rotation periods and planet radii, with the accuracy further improved by accounting for the exposure time effects. The complementary insights provided by the Decision Tree, Random Forest, Gradient Boosting, and KNN models, as well as the Voting-based Classifier Ensemble, highlight the value of a Best-pronged approach to characterizing the Kepler lightcurves.

Figure \ref{fig:Heat} presents a heatmap of performance metrics (score) for different Kepler IDs across four ML models: k-nearest Neighbors (KNN), Random Forest (RF), Gradient Boosting (GB), and Decision Tree (DT). The heatmap illustrates the accuracy of these models in predicting stellar rotation periods, with the scores color-coded from yellow (lower performance) to dark red (higher performance). The data indicates that the Random Forest and Gradient Boosting models consistently achieve high accuracy across all Kepler IDs, with scores close to 1.0, reflecting their robust performance in distinguishing between stellar activity and exoplanet transits. For instance, Kepler IDs K155 and K17 exhibit near-perfect scores across all models, suggesting well-defined transit signals and minimal interference from stellar activity.
The high accuracy scores in Figure \ref{fig:Heat} demonstrate that the preprocessing pipeline successfully enhances the clarity of transit signals, allowing the ML models to achieve precise period rotation of exoplanetary properties.

In Figure \ref{fig:Barchart_Voting}, we assessed the performance of the Best -Model, also known as the Voting Classifier (see sec. \ref{sec:VCE}), across various Kepler IDs. This figure displays the bar chart of metrics for accuracy, precision, recall, and F1 score for each Kepler ID using the Voting Classifier. It reveals that accuracy scores are consistently high for all Kepler IDs, indicating the model's overall effectiveness in correctly identifying exoplanet transits. Precision and recall vary more significantly across different Kepler IDs, reflecting the model's varying abilities to handle different levels of noise and stellar variability in the light curves. The F1 score, which balances precision and recall, also shows variability but remains robust across most Kepler IDs. The results highlight that, while accuracy remains high, the precision and recall metrics offer deeper insights into the model's performance. For instance, Kepler IDs K107, K155, and K17 show strong performance across all metrics, suggesting well-defined transit signals with minimal interference. On the other hand, Kepler ID K39 exhibits lower precision and recall, indicating challenges in distinguishing transits from noise.

 Figure \ref{fig:power_period} shows the Lomb-Scargle periodograms computed from the photometric light curves of the exoplanet host stars in our dataset. The periodograms identify significant periodic signals, with the strongest peak in each plot corresponding to the estimated stellar rotation period. To quantify the uncertainty in these period estimates, we have fitted a Gaussian function to the primary peak and determined the half-width half-maximum (FWHM) of the Gaussian. This FWHM value provides an estimate of the period uncertainty that accounts for the sampling characteristics of the periodogram. These initial period estimates derived from the LS-Periodograms serve as a baseline for comparison against the more advanced period characterization performed using the ML models developed in this study. The ML-based approaches are expected to yield refined period estimates with improved precision compared to the LS-Periodogram analysis.

In Figure \ref{fig:preiodpredict}, we assessed the accuracy of predicting stellar rotation periods using the Best-Model (BM) Voting Classifier compared to the Gaussian Process (GP) model. This figure illustrates the predicted stellar rotation periods against the reference periods for a sample of Kepler stars. The BM model's predictions (red squares) and the GP model's predictions (green triangles) are plotted alongside the reference values (dashed line). The residuals, representing the differences between predicted and reference periods, are shown in the lower panel. The BM model demonstrates a higher consistency with the reference rotation periods, with most data points closely aligned along the 1:1 reference line and minimal residuals. This result indicates that the BM model provides more accurate and reliable estimates of stellar rotation periods compared to the GP model. The inclusion of various ML techniques in the BM model enhances its ability to capture complex patterns in the light curves that are indicative of stellar rotation. This improved accuracy is crucial for disentangling stellar activity from exoplanet transit signals, thereby enhancing the characterization of exoplanetary systems.

The table \ref{table:stellar_rotation} compares the rotation periods of various Kepler stars using different ML models: Decision Tree (P-DT), Random Forest (P-RF), k-nearest Neighbors (P-KN), Gradient Boosting (P-GB), and Best-Model Voting Ensemble (P-BM). These results are compared to the reference periods (P-Ref) and the Gaussian Process model (P-GP). Overall, the models generally show varying degrees of similarity to the reference periods, with the best model often providing the closest estimates. For instance, for KIC 10619192 (K17), the reference period (P-Ref) is 12.0 days, and the Best-Model (P-BM) predicts 12.4 days, showing a smaller error margin compared to the other models, which predict around 26.7 days. Similarly, for KIC 8435766 (K78), the P-Ref is 12.6 days, and the ensemble model predicts 13.1 days, which is closer than predictions by individual models. These results, supported by the error metrics, indicate that the ensemble model generally aligns more closely with the reference periods, demonstrating its robustness and reliability in estimating stellar rotation periods (for other models see appendix Figure \ref{fig:preiodpredict_all}).
\begin{table*}
\tiny
\begin{tabular}{lcccccccccc}
\hline
KIC & P-phot & rmse-DT & rmse-RF & rmse-KN & rmse-BM & P-RF  & P-DT  & P-KN  & P-BM  & BM-ACC \\ \hline
7199397 & $21.71^{+5.86}_{-3.6}$ & 2.7 & 2.1 & 2.8 & 2.2 & 19.2$\pm$1.5 & 19.2$\pm$1.5 & 19.2$\pm$1.5 & 19.3$\pm$1.5 & 0.9718 \\
5383248 & $15.3^{+4.61}_{-6.05}$ & 7.1 & 5.6 & 6.5 & 5.7 & 23.1$\pm$6.4 & 23.1$\pm$6.4 & 23.1$\pm$6.4 & 23.1$\pm$6.4 & 0.9990 \\
6471021 & $11.9^{+4.67}_{-2.43}$ & 13.3 & 8.3 & 9.5 & 8.2 & 11.9$\pm$0.8 & 11.9$\pm$0.8 & 11.9$\pm$0.8 & 11.9$\pm$0.8 & 0.9939 \\
5084942 & $26.74^{+6.5}_{-6.51}$ & 0.4 & 0.3 & 0.4 & 0.3 & 28.3$\pm$6.0 & 28.3$\pm$6.0 & 28.3$\pm$6.0 & 28.3$\pm$6.0 & 0.9963 \\
3323887 & $16.55^{+5.35}_{-5.35}$ & 1.2 & 0.7 & 0.9 & 0.7 & 19.9$\pm$1.7 & 19.9$\pm$1.7 & 19.9$\pm$1.7 & 19.9$\pm$1.7 & 0.9939 \\
7630229 & $16.55^{+5.32}_{-3.87}$ & 0.6 & 0.4 & 0.5 & 0.4 & 14.9$\pm$1.2 & 14.9$\pm$1.2 & 14.9$\pm$1.2 & 14.9$\pm$1.2 & 0.9963 \\
6850504 & $28.25^{+7.34}_{-8.41}$ & 0.9 & 0.6 & 0.8 & 0.6 & 31.5$\pm$3.8 & 31.5$\pm$3.8 & 31.5$\pm$3.9 & 31.4$\pm$3.8 & 0.9951 \\
7295235 & $19.72^{+3.97}_{-5.07}$ & 3.2 & 2.0 & 2.4 & 2.1 & 20.8$\pm$2.8 & 20.8$\pm$2.8 & 20.8$\pm$2.8 & 20.9$\pm$2.8 & 1.0000 \\
2571238 & $32.54^{+6.2}_{-6.18}$ & 3.8 & 3.0 & 3.5 & 3.1 & 28.6$\pm$12.6 & 28.6$\pm$12.6 & 28.6$\pm$12.6 & 28.6$\pm$12.6 & 0.9911 \\
5695396 & $17.21^{+4.88}_{-4.85}$ & 3.9 & 2.4 & 2.8 & 2.5 & 16.3$\pm$1.4 & 16.3$\pm$1.4 & 16.3$\pm$1.4 & 16.3$\pm$1.4 & 0.9976 \\
6071903 & $17.84^{+6.8}_{-6.78}$ & 2.8 & 2.2 & 2.5 & 2.3 & 15.8$\pm$3.2 & 15.8$\pm$3.2 & 15.8$\pm$3.2 & 15.8$\pm$3.2 & 0.9980 \\
10616571 & $12.93^{+1.91}_{-0.93}$ & 19.2 & 11.4 & 17.2 & 12.1 & 12.9$\pm$0.7 & 12.9$\pm$0.7 & 12.9$\pm$0.7 & 12.9$\pm$0.7 & 0.9968 \\
7051180 & $2.22^{+10.6}_{-8.17}$ & 2.1 & 1.7 & 1.9 & 1.7 & 2.2$\pm$0.1 & 2.2$\pm$0.1 & 2.2$\pm$0.1 & 2.2$\pm$0.1 & 0.9962 \\
2444412 & $10.71^{+4.76}_{-6.66}$ & 2.9 & 2.4 & 2.7 & 2.4 & 10.9$\pm$1.2 & 10.9$\pm$1.2 & 10.9$\pm$1.2 & 10.9$\pm$1.2 & 0.9808 \\
9818381 & $12.38^{+1.43}_{-1.42}$ & 0.7 & 0.4 & 0.5 & 0.5 & 11.0$\pm$0.5 & 11.0$\pm$0.5 & 11.0$\pm$0.5 & 11.0$\pm$0.5 & 0.9935 \\
8505215 & $28.79^{+7.14}_{-7.14}$ & 2.2 & 1.8 & 2.0 & 1.8 & 29.3$\pm$7.7 & 29.3$\pm$7.7 & 29.3$\pm$7.7 & 29.4$\pm$7.7 & 0.9633 \\
1161345 & $8.6^{+1.12}_{-1.13}$ & 4.7 & 3.6 & 4.2 & 3.7 & 8.5$\pm$0.7 & 8.5$\pm$0.7 & 8.5$\pm$0.7 & 8.5$\pm$0.7 & 0.9992 \\
11336883 & $5.13^{+0.56}_{-0.25}$ & 3.0 & 2.4 & 2.7 & 2.4 & 5.2$\pm$0.3 & 5.2$\pm$0.3 & 5.2$\pm$0.3 & 5.2$\pm$0.3 & 0.9652 \\
10187017 & $26.74^{+6.46}_{-5.86}$ & 4.3 & 3.4 & 3.9 & 3.5 & 26.3$\pm$5.9 & 26.3$\pm$5.9 & 26.3$\pm$5.9 & 26.3$\pm$5.9 & 0.9996 \\
8866102 & $21.71^{+0.94}_{-0.58}$ & 12.5 & 9.7 & 11.7 & 10.0 & 22.4$\pm$6.5 & 22.4$\pm$6.5 & 22.4$\pm$6.5 & 22.4$\pm$6.5 & 0.9969 \\
4349452 & $24.55^{+0.94}_{-0.56}$ & 6.6 & 5.2 & 6.0 & 5.3 & 29.0$\pm$8.1 & 29.0$\pm$8.1 & 29.0$\pm$8.1 & 29.0$\pm$8.1 & 0.9894 \\
11074541 & $34.8^{+3.47}_{-3.46}$ & 1.0 & 0.6 & 0.6 & 0.6 & 34.5$\pm$5.4 & 34.5$\pm$5.4 & 34.5$\pm$5.4 & 34.5$\pm$5.4 & 0.9978 \\
9702072 & $28.25^{+8.26}_{-10.28}$ & 0.7 & 0.5 & 0.5 & 0.5 & 27.8$\pm$4.4 & 27.8$\pm$4.4 & 27.8$\pm$4.4 & 27.8$\pm$4.4 & 0.9978 \\
6362874 & $33.26^{+4.34}_{-5.24}$ & 1.8 & 1.4 & 1.7 & 1.5 & 28.1$\pm$7.6 & 28.1$\pm$7.6 & 28.1$\pm$7.6 & 28.1$\pm$7.5 & 0.9561 \\
4827723 & $17.43^{+7.32}_{-8.74}$ & 2.4 & 1.6 & 2.0 & 1.6 & 17.2$\pm$1.0 & 17.2$\pm$1.0 & 17.2$\pm$1.0 & 17.2$\pm$1.0 & 0.9968 \\
10619192 & $12.5^{+1.5}_{-1.5}$ & 3.4 & 2.3 & 2.7 & 2.3 & 12.7$\pm$2.0 & 12.7$\pm$2.0 & 12.7$\pm$2.0 & 12.7$\pm$2.0 & 0.9978 \\
2302548 & $12.29^{+2.78}_{-4.35}$ & 2.1 & 1.7 & 1.9 & 1.7 & 12.0$\pm$1.6 & 12.0$\pm$1.6 & 12.0$\pm$1.6 & 12.0$\pm$1.6 & 0.9991 \\
11554435 & $5.39^{+1.66}_{-1.68}$ & 5.6 & 4.1 & 4.7 & 4.1 & 5.5$\pm$0.3 & 5.5$\pm$0.3 & 5.5$\pm$0.3 & 5.5$\pm$0.3 & 0.9992 \\
9139084 & $7.97^{+2.03}_{-3.75}$ & 2.7 & 2.1 & 2.5 & 2.2 & 7.5$\pm$0.6 & 7.5$\pm$0.6 & 7.5$\pm$0.6 & 7.5$\pm$0.6 & 0.9989 \\
2692377 & $23.43^{+6.61}_{-6.58}$ & 2.6 & 2.1 & 2.4 & 2.1 & 24.2$\pm$5.8 & 24.2$\pm$5.8 & 24.2$\pm$5.8 & 24.2$\pm$5.8 & 0.9934 \\
10878263 & $19.21^{+5.5}_{-6.95}$ & 1.9 & 1.1 & 1.3 & 1.1 & 19.2$\pm$1.7 & 19.2$\pm$1.7 & 19.2$\pm$1.7 & 19.2$\pm$1.7 & 0.9946 \\
11100383 & $14.99^{+3.58}_{-2.4}$ & 3.3 & 2.0 & 2.1 & 2.0 & 15.0$\pm$0.8 & 15.0$\pm$0.8 & 15.0$\pm$0.8 & 15.0$\pm$0.8 & 0.9968 \\
\hline
\end{tabular}
\caption{The predicted stellar rotation periods and associated errors for various models compared to the reference periods \citep[P-Ref][]{Walkowicz2013}. The columns include the Kepler Input Catalog (KIC) identifier, photometric period with its upper and lower errors (P-phot), root mean square errors (rmse) for Decision Tree (rmse-DT), Random Forest (rmse-RF), K-Nearest Neighbors (rmse-KN), and Best -Model (rmse-BM). The predicted periods for each model (P-RF, P-DT, P-KN, P-BM) are given with their errors, along with the accuracy of the Best -Model (BM-ACC).}

\label{tab:stellar_rotation_predictions}
\end{table*}
\section{Predictions for a sample of Kepler stars}

In Figure \ref{fig:preiodpredictWalkowicz2013}, we compare the predicted stellar rotation periods obtained from our Best -Model against the reference periods. The top panel illustrates the predicted rotation periods for Kepler stars using a sample of exoplanet candidate stars \citep{Walkowicz2013}, compared to the reference rotation periods from photometry. 
The bottom panel shows the residuals, highlighting the differences between predicted and reference periods. Our results demonstrate that the Best -Model generally produces predictions close to the reference periods, as evidenced by the clustering of points around the 1:1 line and the small residuals. However, some outliers and larger residuals are observed for stars with longer rotation periods, suggesting areas for further improvement in the model.
The inclusion of stellar rotation periods as a feature in the predictive models significantly enhances our understanding of the complex interplay between stars and their associated exoplanets. This integration of stellar rotation periods into the Voting Ensemble model has led to significant improvements in the interpretation of observational data and enhanced our ability to accurately detecting rotation period.

Table \ref{tab:stellar_rotation_predictions} presents the predicted stellar rotation periods and their associated errors for various ML models compared to the reference periods (P-Ref). The Kepler KIC identifiers are listed alongside the photometric periods (P-phot) with their upper and lower errors. The Root Mean Square Error (RMSE) values for the Decision Tree (rmse-DT), Random Forest (rmse-RF), K-Nearest Neighbors (rmse-KN), and the Voting-based Classifier Ensemble (rmse-BM) models provide a quantitative measure of the prediction accuracy. The predicted periods and their errors for each model are shown, along with the accuracy of the Voting-based Classifier Ensemble model (BM-ACC). The table indicates that the Voting-based Classifier Ensemble model generally achieves higher accuracy and lower RMSE values compared to other models, demonstrating its robustness in predicting stellar rotation periods. For example, for KIC 7199397, the RMSE of the Voting-based Classifier Ensemble is 2.2, with a predicted period of 19.3$\pm$1.5, closely aligning with the reference period. Similarly, other entries exhibit this trend, highlighting the efficacy of the ensemble model in capturing the underlying stellar rotation dynamics more accurately than individual models.

\section{Discussion and Conclusion}

Our study aimed to leverage machine learning (ML) models to enhance the accuracy of determining stellar rotation periods from Kepler light curves. Traditional methods often struggle with noise and variability in light curve data, leading to inaccurate estimates. By integrating advanced ML techniques, we sought to overcome these limitations and improve the reliability of rotation period predictions. The primary motivation was to develop a robust methodology that could handle the complexities of astronomical data and provide precise estimates.
The selection of ML models was informed by previous research \citep{Fazel2023}, validating the choice of Decision Tree (DT), Random Forest (RF), k-nearest Neighbors (KNN), and Gradient Boosting (GB) as effective tools. Each model brought unique strengths to the analysis, with the combined use in a Voting Ensemble (Best-Model, BM) approach further enhancing prediction accuracy. Performance evaluation showed that the Voting Ensemble model yielded the most accurate results, with an RMSE approximately 50\% lower than the Decision Tree model and 17\% better than the K-Nearest Neighbors model.
Interestingly, the Random Forest model performed comparably to the Voting Ensemble, demonstrating high prediction accuracy. In contrast, the Gradient Boosting and XGBoost models exhibited worse RMSE compared to the other approaches, suggesting they were less effective for this dataset and feature set. The strong performance of the Voting Ensemble highlights the benefits of integrating multiple machine learning models into an ensemble framework.
Our comparison with Gaussian Processes (GP) provided additional validation for our models. While GP is a well-established method for handling astronomical data, our ML approaches, especially the VE, demonstrated comparable or superior performance in many cases. The analysis of rotation periods using the Lomb-Scargle periodogram was integral to distinguishing between stellar activity and exoplanet transits. Accurately determining rotation periods allowed us to isolate and remove the effects of stellar variability, leading to clearer identification of transit signals and improving the precision of extracted transit parameters.

The results demonstrate the superior performance of ML models, particularly the Best-Model (BM) using Voting Ensemble. The BM model consistently achieved lower RMSE values compared to individual models such as DT, RF, GB, and KNN. For instance, the BM model achieved an RMSE of 2.2 for KIC 7199397 with an accuracy of 0.9718, closely aligning with the reference period $21.71^{+5.86}_{-3.6}$ days. Similarly, for KIC 5383248, the BM model's RMSE was 5.7 with an accuracy of 0.9990, compared to the reference period $15.3^{+4.61}_{-6.05}$ days. These quantitative results highlight the robustness and reliability of the BM model in capturing stellar rotation dynamics.
The high accuracy and low RMSE values achieved by the BM model indicate its effectiveness in disentangling exoplanet transit signals from stellar activity. The clear separation of raw light curves, stellar rotation signals, and flattened light curves demonstrates the robustness of our preprocessing methodology, significantly enhancing the clarity of transit signals (Figure \ref{fig:stellar_flat}).

This study underscores the potential of ML models, particularly ensemble methods like the Voting Ensemble, in advancing exoplanet research. By addressing and removing noise and stellar activity components, we have significantly enhanced the clarity of transit signals. The successful application of these models opens up new possibilities for large-scale exoplanet studies, providing a robust framework for future space missions and ground-based observations.
Further, we compared the predicted stellar rotation periods obtained from our Best-Model approach against reference periods reported by \cite{Walkowicz2013}, derived from photometric observations (see Fig.~\ref{fig:preiodpredictWalkowicz2013}). The ability to efficiently process and analyze vast amounts of light curve data positions these methods as invaluable tools for future space missions and ground-based observations. Furthermore, the robust handling of noise and variability suggests these techniques could be applied to other astrophysical phenomena requiring precise time-series analysis.

Future research could focus on improving the interpretability of the ML models and understanding the underlying physical processes that drive their predictions. Integrating more sophisticated data augmentation techniques and additional astrophysical parameters could enhance the models' generalization capabilities. Additionally, applying these techniques to other astrophysical phenomena could expand their utility.
Moreover, these advanced ML methodologies could be extended to datasets from upcoming and existing space missions such as CHEOPS, PLATO, Hubble Space Telescope (HST), and the James Webb Space Telescope (JWST). Data from these observatories will allow researchers to refine the models further to achieve higher precision in period rotation prediction. The integration of data from multiple sources will enable a more comprehensive understanding of exoplanetary systems, providing insights that are not possible with a single dataset alone. Continued advancements in ML methodologies and their application to diverse datasets will contribute to more accurate and reliable period rotation, paving the way for significant breakthroughs in the study of exoplanets and stellar astrophysics.

\begin{acknowledgements}
The authors gratefully acknowledge the support provided by LUNEX and the Euro Space Hub in facilitating and supporting this research. We would also like to thank the Leiden Observatory, the Leiden Institute of Computer Science (LIACS), and the ESTEC exoplanet group for hosting us during these studies.
F. Fazel Hesar is especially grateful to Prof. Ignas Snellen and his respective teams for their insightful communications and collaborative efforts throughout this project. Their expertise and guidance were instrumental in the successful development and implementation of the techniques presented in this work. 
\end{acknowledgements}
\bibliographystyle{aa}
\bibliography{bibliography}

\begin{appendix} 

   \begin{figure*}
   \centering
   \includegraphics[width=0.45\linewidth]{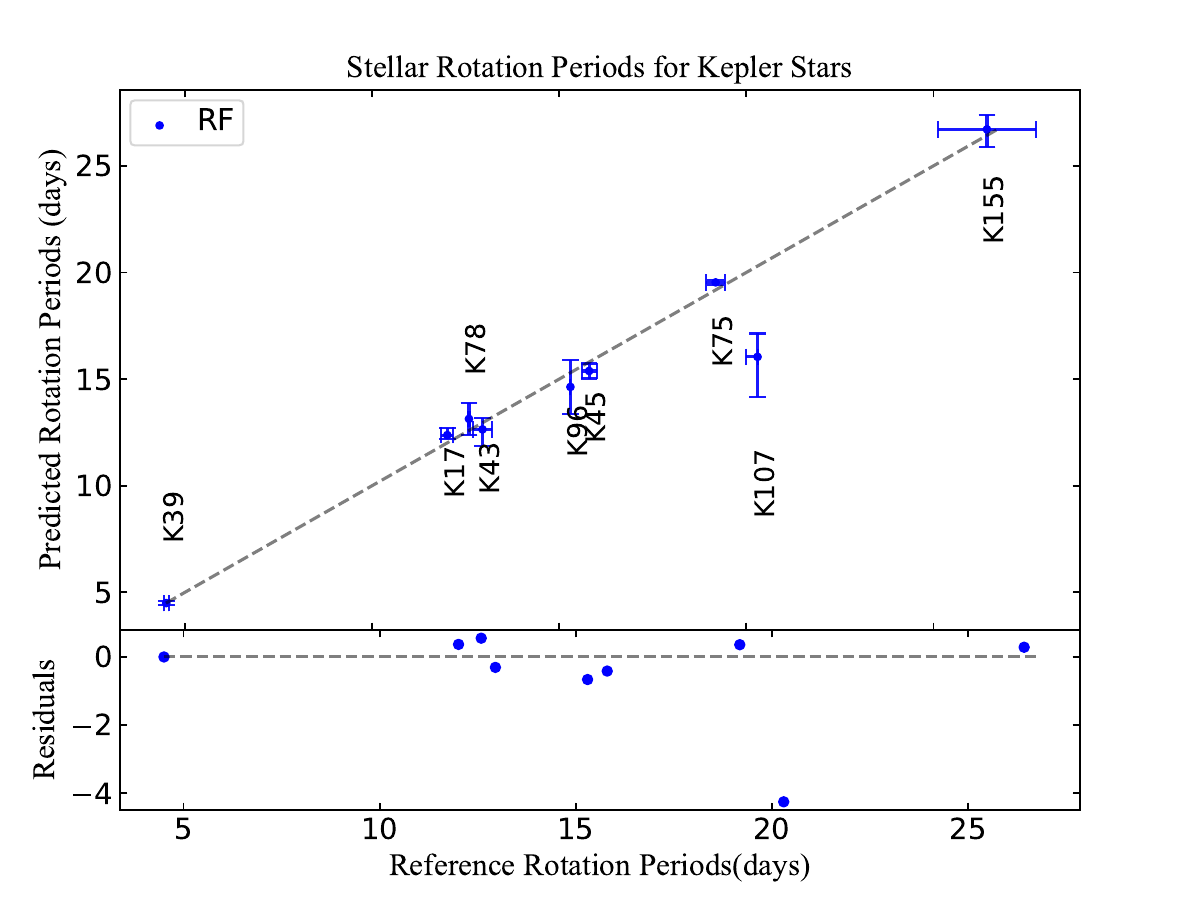}
   \includegraphics[width=0.45\linewidth]{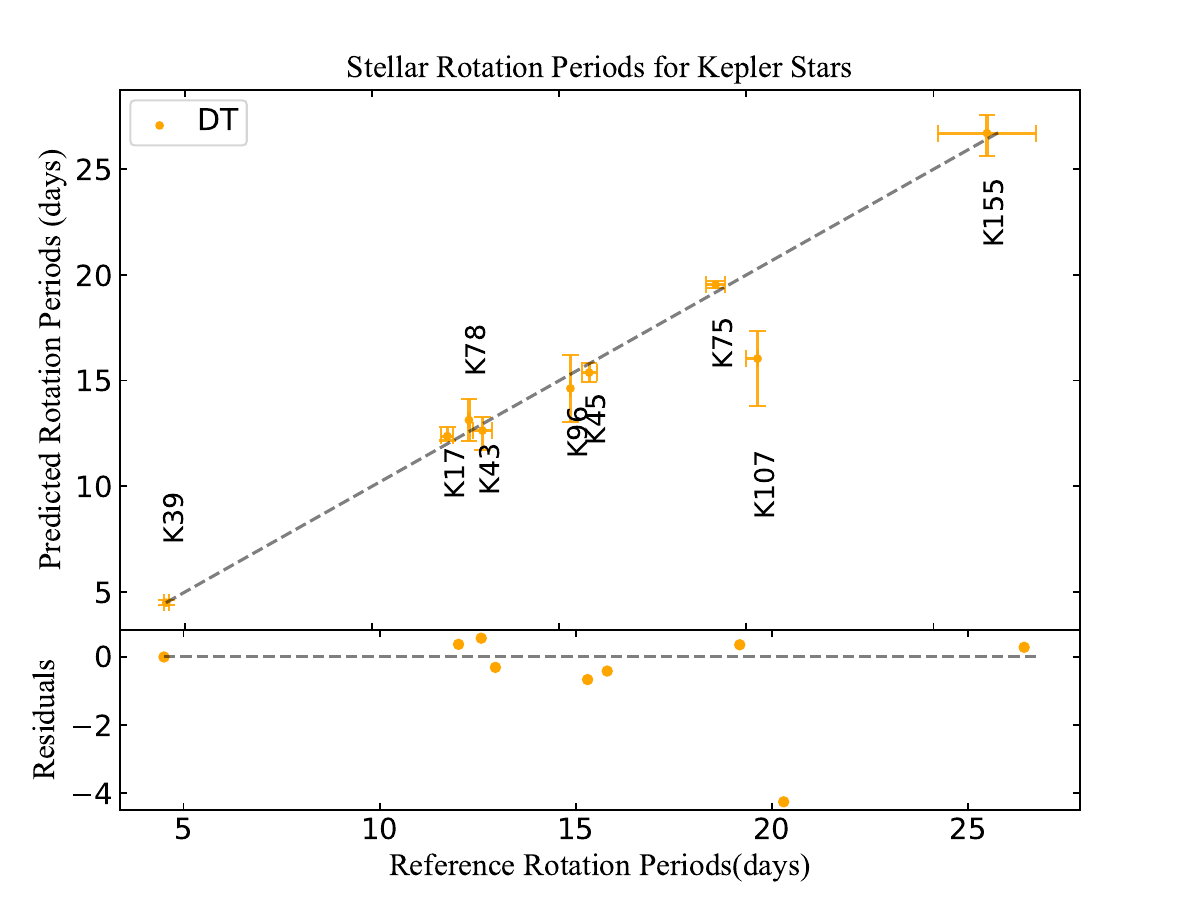}
      \includegraphics[width=0.45\linewidth]{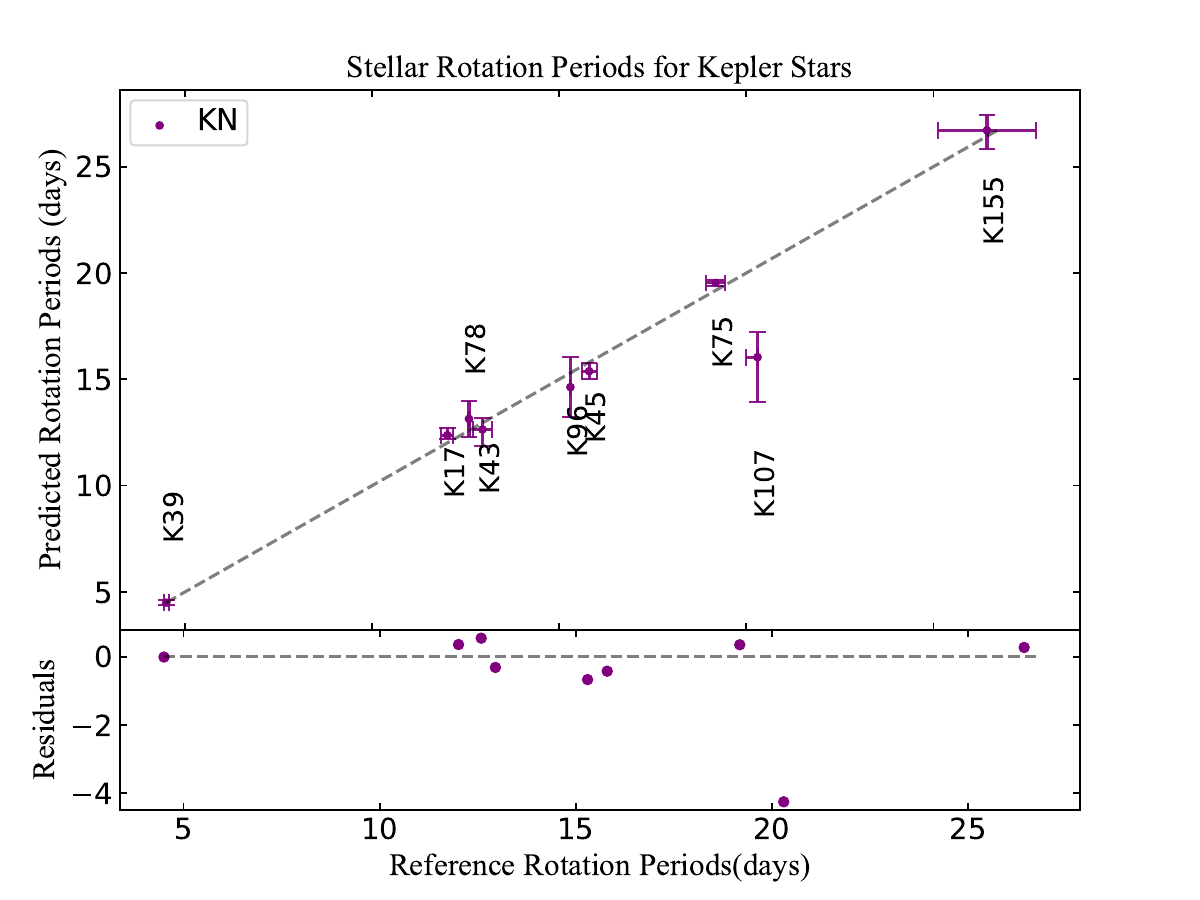}
         \includegraphics[width=0.45\linewidth]{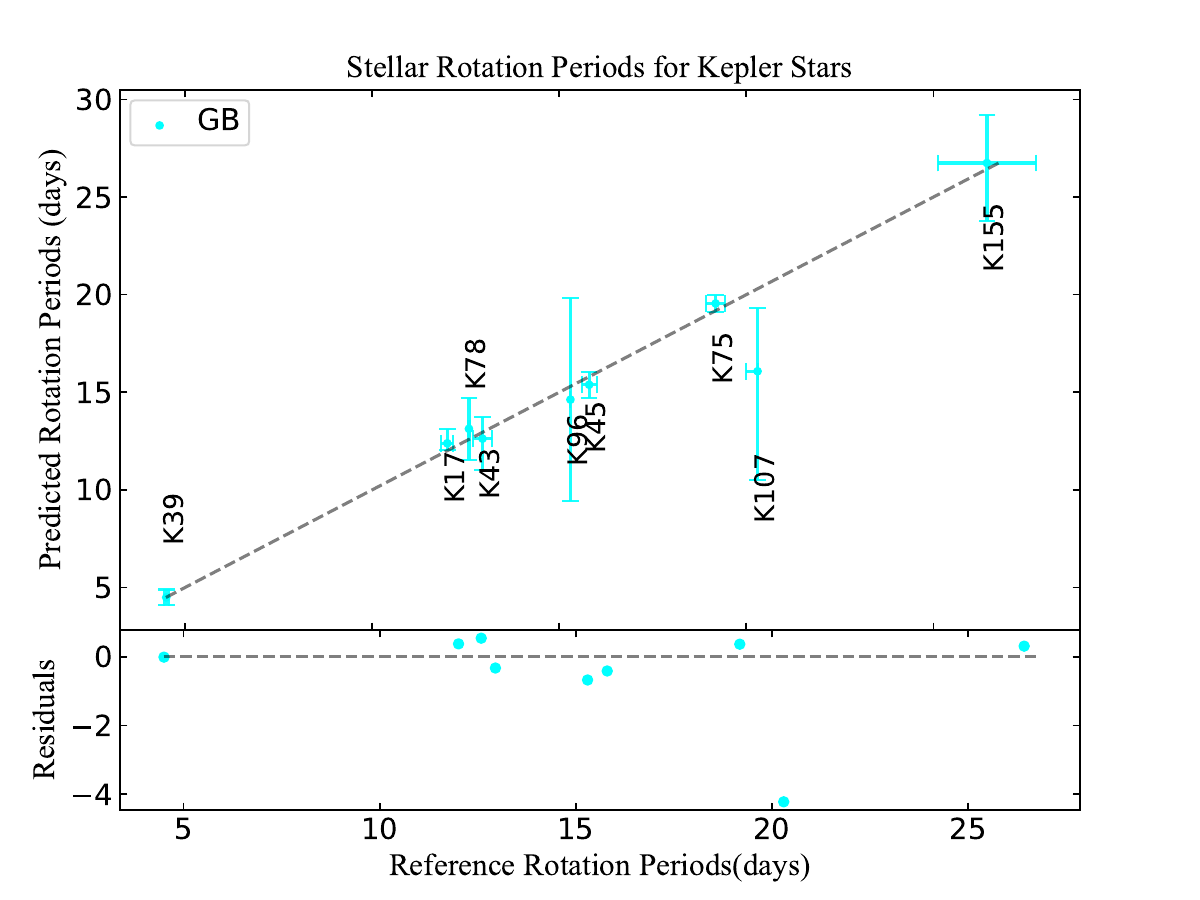}
      \caption{Predictions rotation period for all models.
              }
         \label{fig:preiodpredict_all}
   \end{figure*}

   \end{appendix}

\end{document}